\documentstyle[12pt]{article}

\catcode`\@=11
\long\def\@makefntext#1{
\protect\noindent \hbox to 3.2pt {\hskip-.9pt  
$^{{\ninerm\@thefnmark}}$\hfil}#1\hfill}                

\def\@makefnmark{\hbox to 0pt{$^{\@thefnmark}$\hss}}  
        
\def\ps@myheadings{\let\@mkboth\@gobbletwo
\def\@oddhead{\hbox{}
\rightmark\hfil\ninerm\thepage}   
\def\@oddfoot{}\def\@evenhead{\ninerm\thepage\hfil
\leftmark\hbox{}}\def\@evenfoot{}
\def\sectionmark##1{}\def\subsectionmark##1{}}

\renewcommand{\thefootnote}{\fnsymbol{footnote}}

\newcounter{sectionc}\newcounter{subsectionc}\newcounter{subsubsectionc}
\renewcommand{\section}[1] {\vspace*{0.6cm}\addtocounter{sectionc}{1} 
\setcounter{subsectionc}{0}\setcounter{subsubsectionc}{0}\noindent 
        {\normalsize\bf\thesectionc. #1}\par\vspace*{0.4cm}}
\renewcommand{\subsection}[1] {\vspace*{0.6cm}\addtocounter{subsectionc}{1} 
        \setcounter{subsubsectionc}{0}\noindent 
        {\normalsize\it\thesectionc.\thesubsectionc. #1}\par\vspace*{0.4cm}}
\renewcommand{\subsubsection}[1]
{\vspace*{0.6cm}\addtocounter{subsubsectionc}{1}
        \noindent {\normalsize\rm\thesectionc.\thesubsectionc.\thesubsubsectionc. 
        #1}\par\vspace*{0.4cm}}

\newcounter{appendixc}
\newcounter{subappendixc}[appendixc]
\newcounter{subsubappendixc}[subappendixc]

\renewcommand{\appendix}[1] {\vspace*{0.6cm}
        \refstepcounter{appendixc}
        \setcounter{figure}{0}
        \setcounter{table}{0}
        \setcounter{equation}{0}
        \renewcommand{\thefigure}{\Alph{appendixc}.\arabic{figure}}
        \renewcommand{\thetable}{\Alph{appendixc}.\arabic{table}}
        \renewcommand{\theappendixc}{\Alph{appendixc}}
        \renewcommand{\theequation}{\Alph{appendixc}.\arabic{equation}}
        \noindent{\bf Appendix \theappendixc #1}\par\vspace*{0.4cm}}

\def\abstracts#1{{
        \centering{\begin{minipage}{12.2truecm}\footnotesize\baselineskip=12pt\noindent
        \centerline{\footnotesize ABSTRACT}\vspace*{0.3cm}
        \parindent=0pt #1
        \end{minipage}}\par}} 

\renewenvironment{thebibliography}[1]
        {\begin{list}{\arabic{enumi}.}
        {\usecounter{enumi}\setlength{\parsep}{0pt}
\setlength{\leftmargin 1.25cm}{\rightmargin 0pt}
         \setlength{\itemsep}{0pt} \settowidth
        {\labelwidth}{#1.}\sloppy}}{\end{list}}

\newcounter{itemlistc}
\newcounter{romanlistc}
\newcounter{alphlistc}
\newcounter{arabiclistc}

\newcommand{\fcaption}[1]{
        \refstepcounter{figure}
        \setbox\@tempboxa = \hbox{\footnotesize Fig.~\thefigure. #1}
        \ifdim \wd\@tempboxa > 6in
           {\begin{center}
        \parbox{6in}{\footnotesize\baselineskip=12pt Fig.~\thefigure. #1}
            \end{center}}
        \else
             {\begin{center}
             {\footnotesize Fig.~\thefigure. #1}
              \end{center}}
        \fi}

\newcommand{\tcaption}[1]{
        \refstepcounter{table}
        \setbox\@tempboxa = \hbox{\footnotesize Table~\thetable. #1}
        \ifdim \wd\@tempboxa > 6in
           {\begin{center}
        \parbox{6in}{\footnotesize\baselineskip=12pt Table~\thetable. #1}
            \end{center}}
        \else
             {\begin{center}
             {\footnotesize Table~\thetable. #1}
              \end{center}}
        \fi}

\def\@citex[#1]#2{\if@filesw\immediate\write\@auxout
        {\string\citation{#2}}\fi
\def\@citea{}\@cite{\@for\@citeb:=#2\do
        {\@citea\def\@citea{,}\@ifundefined
        {b@\@citeb}{{\bf ?}\@warning
        {Citation `\@citeb' on page \thepage \space undefined}}
        {\csname b@\@citeb\endcsname}}}{#1}}

\newif\if@cghi
\def\cite{\@cghitrue\@ifnextchar [{\@tempswatrue
        \@citex}{\@tempswafalse\@citex[]}}
\def\citelow{\@cghifalse\@ifnextchar [{\@tempswatrue
        \@citex}{\@tempswafalse\@citex[]}}
\def\@cite#1#2{{$\null^{#1}$\if@tempswa\typeout
        {IJCGA warning: optional citation argument 
        ignored: `#2'} \fi}}

 1
 1
 1

\font\ninerm=cmr9

\def\be{\begin{equation}}
\def\ee{\end{equation}}
\def\bea{\begin{eqnarray}}
\def\eea{\end{eqnarray}}
\def\bean{\begin{eqnarray*}}
\def\eean{\end{eqnarray*}}
\def\ba{\begin{eqnarray}} \def\ea{\end{eqnarray}}
\def\stareq{\ {\buildrel{*}\over =}\ }

\newcommand{\baz}{\overline{0}}

\newcommand{\bai}{\overline{i}}

\def\6{\partial} \def\a{\alpha} \def\b{\beta}
\def\g{\gamma} \def\d{\delta} \def\ve{\varepsilon} 
 \def\h{\eta} \def\th{\theta}
\def\vt{\vartheta} \def\k{\kappa} 

 \def\ph{\varphi} 
 \def\G{\Gamma} 
  
 \def\Ps{\Psi}

\def\rR{{\buildrel {\{\}} \over R}}
\def\rG{{\buildrel {\{\}} \over \G}}
\def\rE{{\buildrel {\{\}} \over G}}
\def\={\!\!\!&=&\!\!\!}
\def\+{\!\!\!&&\!\!\!+~}
\def\-{\!\!\!&&\!\!\!-~}

\def\lr{\Longrightarrow}
\def\stareq{\buildrel\ast\over =}
\def\chris#1#2#3{#1\brace #2 #3}

\newcommand{\RR}{{\cal R}}
\newcommand{\TT}{{\cal T}}

\def\vta{\vartheta}
\def\a{\alpha}
\def\b{\beta}
\def\d{\delta}
\def\g{\gamma}

\def\hodge {{}^*\!}

\def\ltextindent#1{\hbox to \hangindent{#1\hss}\ignorespaces}

\def\sqr#1#2{{\vcenter{\hrule height.#2pt\hbox{\vrule width.#2pt
height#1pt \kern#1pt \vrule width.#2pt}\hrule height.#2pt}}}

\def\negenspace{\kern-1.1em}
\def\quer{\negenspace\nearrow}
\def\negenspaceexp{\kern-0.5em}

\def\semidirect{\;{\rlap{$\subset$}\times}\;}

\def\semidirect{\;{\rlap{$\supset$}\times}\;}
\input{epsf.tex}
\def\mbold#1{\hbox{\boldmath $#1$}}
\begin{document}
\bigskip

\centerline{\normalsize\bf On the Gauge Aspects of Gravity${}^\dagger$}

\bigskip
\bigskip
\centerline{\footnotesize Frank Gronwald}
\vspace*{0.3cm}
\centerline{\footnotesize and}
\vspace*{0.3cm}
\centerline{\footnotesize Friedrich W. Hehl}
\baselineskip=13pt
\medskip
\centerline{\footnotesize\it Institute for Theoretical Physics, University
                             of Cologne}
\centerline{\footnotesize\it D-50923 K{\"o}ln, Germany}
\centerline{\footnotesize E-mail: fg@thp.uni-koeln.de, hehl@thp.uni-koeln.de}

\vspace*{0.9cm} \abstracts{
We give a short outline, in Sec.\ 2, of the historical development 
of the gauge idea as applied to internal ($U(1),\, SU(2),\dots$) 
and external ($R^4,\,SO(1,3),\dots$) symmetries and stress the 
fundamental importance of the corresponding conserved currents. 
In Sec.\ 3, experimental results with neutron interferometers in the 
gravitational field of the earth, as interpreted by means of the 
equivalence principle, can be predicted by means of the Dirac 
equation in an accelerated and rotating reference frame.  Using 
the Dirac equation in such a non-inertial frame, we describe how 
in a gauge-theoretical approach (see Table 1) the 
Einstein-Cartan theory, residing in a Riemann-Cartan spacetime 
encompassing torsion and curvature, arises as the simplest 
gravitational theory. This is set in contrast to the Einsteinian 
approach yielding general relativity in a Riemannian spacetime. 
In Secs.\ 4 and 5 we consider the conserved energy-momentum 
current of matter and gauge the associated translation subgroup. 
The Einsteinian teleparallelism theory which emerges is shown to be 
equivalent, for spinless matter and for electromagnetism, to 
general relativity. Having successfully gauged the translations, 
it is straightforward to gauge the four-dimensional affine 
group $R^4\semidirect GL(4,R)$ or its Poincar\'e subgroup 
$R^4\semidirect SO(1,3)$. We briefly report on these results
in Sec.\ 6 (metric-affine geometry) and in Sec.\ 7 (metric-affine
field equations (\ref{zeroth}, \ref{first}, \ref{second})). 
Finally, in Sec.\ 8, 
we collect some models, 
currently under discussion, which bring life into the 
metric-affine gauge framework developed.}

\vspace*{3cm} \setcounter{footnote}{-1} \footnote{${}^\dagger$ Proc.\ 
  of the 14th Course of the School of Cosmology and Gravitation on
  {\it Quantum Gravity}, held at Erice, Italy, May 1995, P.G.\
  Bergmann, V.\ de Sabbata, and H.-J.\ Treder, eds. (World Scientific,
  Singapore 1996) to be published.}

\pagebreak

\noindent
\centerline{\bf Contents}
\medskip
\begin{description}
\item{$\,\;$1.} Introduction
\item{$\,\;$2.} Remarks on the history of the gauge idea
\subitem{2.1.} General relativity and Weyl's $U(1)$-gauge theory
\subitem{2.2.} Yang-Mills and the structure of a gauge theory
\subitem{2.3.} Gravity and the Utiyama-Sciama-Kibble approach
\subitem{2.4.} E.\ Cartan's analysis of general relativity and its 
consequences
\item{$\,\;$3.} Einstein's and the gauge approach to gravity
\subitem{3.1.} Neutron matter waves in the gravitational field
\subitem{3.2.} Accelerated and rotating reference frame
\subitem{3.3.} Dirac matter waves in a non-inertial frame of reference
\subitem{3.4.} `Deriving' a theory of gravity: Einstein's method as opposed
               to the 
\subitem{$\qquad\;$}gauge procedure
\item{$\,\;$4.} Conserved momentum current, the heuristics of the translation 
          gauge 
\subitem{4.1.} Motivation
\subitem{4.2.} Active and passive translations
\subitem{4.3.} Heuristic scheme of translational gauging
\item{$\,\;$5.} Theory of the translation gauge: 
{}From Einsteinian teleparallelism to GR
\subitem{5.1.} Translation gauge potential
\subitem{5.2.} Lagrangian
\subitem{5.3.} Transition to GR
\item{$\,\;$6.} Gauging of the affine group $R^{4}\semidirect GL(4,R)$
\item{$\,\;$7.} Field equations of metric-affine gauge theory (MAG)
\item{$\,\;$8.} Model building: Einstein-Cartan theory and beyond
\subitem{8.1.} Einstein-Cartan theory EC
\subitem{8.2.} Poincar\'e gauge theory PG, the quadratic version
\subitem{8.3.} Coupling to a scalar field
\subitem{8.4.} Metric-affine gauge theory MAG
\item{$\,\;$9.} Acknowledgments
\item{10.} References  
\end{description}

\pagebreak

\bigskip \bigskip \parbox[t]{13cm}{ From a letter of A.\ Einstein to
  F.\ Klein of 1917 March 4 (translation)\cite{Pais}: \\ 

{\it ``\dots Newton's theory \dots represents the gravitation\-al
field in a seemingly complete way by means of the potential
$\Phi$. This description proves to be wanting; the functions
$g_{\mu\nu}$ take its place. But I do not doubt that the day will 
come when that de\-scription, too, will have to yield to another
one, for reasons which at present we do not yet surmise.
I believe that this process of deepening the theory has 
no limits \dots " \it} \\ }

\normalsize\baselineskip=15pt
\setcounter{footnote}{0}
\renewcommand{\thefootnote}{\alph{footnote}}
\section{Introduction}
\renewcommand{\thefootnote}{\alph{footnote}}

\noindent$\bullet$ What can we learn if we look at gravity and, more
specifically, at general relativity theory (GR) from the point of view
of classical gauge field theory?  This is the question underlying our
present considerations.  The answer

\medskip
\noindent$\bullet$ leads to a better understanding of the
interrelationship between the {\it metric} and {\it affine} properties
of spacetime and of the {\it group} structure related to gravity.
Furthermore, it

\medskip \noindent$\bullet$ suggests certain classical
field-theoretical generalizations of Einstein's theory, such as
Einstein--Cartan theory, Einsteinian teleparallelism theory,
Poincar\'e gauge theory, Metric-Affine Gravity, that is, it leads to a
deepening of the insight won by GR.

\medskip We recently published a fairly technical review article on
our results\cite{PRs}. These lectures can be regarded as a
down-to-earth introduction into that subject. We refrain from citing
too many articles since we gave an overview\footnote{In the meantime
  we became aware that in our Physics Reports we should have cited
  additionally the work of Mistura\cite{Mistura} on the physical
  interpretation of torsion, of Pascual-S\'anchez\cite{PS1,PS2,PS3} on
  teleparallelism, inter alia, of Perlick\cite{Perlick} on observers
  in Weyl spacetimes, and that of Ponomariov and Obukhov\cite{Pono} on
  metric-affine spacetimes and gravitational theories with quadratic
  Lagrangians.} $\,$ of the existing literature in  ref.(\cite{PRs}).

\section{Remarks on the history of the gauge idea}
\subsection{General relativity and Weyl's $U(1)$-gauge theory}

Soon after Einstein in 1915/16 had proposed his gravitational theory,
namely general relativity (GR), Weyl extended it in 1918 in order to
include -- besides gravitation~-- electromagnetism in a unified way.
Weyl's theoretical concept was that of recalibration or {\it gauge
  invariance} of length. In Weyl's opinion, the integrability of
length in GR is a remnant of an era dominated by action-at-a-distance
theories which should be abandoned. In other words, if in GR we
displace a meter stick from one point of spacetime to another one, it
keeps its length, i.e., it can be used as a standard of length
throughout spacetime; an analogous argument is valid for a clock. In
contrast, Weyl's unified theory of gravitation and electromagnetism of
1918 is set up in such a way that the unified Lagrangian is invariant
under recalibration or re-gauging.
\par
For that purpose, Weyl extended the geometry of spacetime from the
(pseudo-) Riemannian geometry with its Levi-Civita connection 
$\Gamma^{\{\}}_{\alpha\beta}$ to a Weyl
space with an additional (Weyl) covector field
$Q=Q_\alpha\,\vartheta^\alpha$, where $\vartheta^\alpha$ denotes the
field of coframes of the underlying four-dimensional differentiable
manifold. The Weyl connection one-form reads 
\begin{equation}
\Gamma^{\rm W}_{\alpha\beta}=\Gamma^{\{\}}_{\alpha\beta}+ {1\over 2}
(g_{\alpha\beta}\,Q-\vartheta_\alpha\,Q_\beta+\vartheta_\beta\,Q_\alpha)
\,.\label{GammaW} \end{equation}
The additional freedom of having a new one-form (or covector) field
$Q$ at one's disposal was used by Weyl in order to accommodate
Maxwell's field. He identified the electromagnetic potential $A$ with
$Q$.
\par 
Weyl's theory turned out to be non-viable, at least in the sense and
on the level of ordinary length and time measurement. However, his
concept of gauge invariance survived in the following way: When
quantum mechanics was developed, it became clear that (in the
Schr\"odinger representation) the wave function $\Psi$ of an electron,
for example, is only determined up to an arbitrary phase $\phi$: \be
\Psi \longrightarrow e^{i\phi}\Psi \,,\qquad
\phi=const\,.\label{rigid}\ee The set of all phase transformations
builds up the one-dimensional Abelian Lie group $U(1)$ of unitary
transformations. If one substitutes, according to (\ref{rigid}), the
wave function in the Dirac equation by a phase transformed wave
function, no observables will change; they are invariant under
`rigid', i.e., constant phase transformations. This is an elementary
fact of quantum mechanics.
\par
In 1929, Weyl revitalized his gauge idea: Isn't it against the spirit
of field theory to implement a rigid phase transformation
(\ref{rigid}) `at once' all over spacetime, he asked. Shouldn't we
postulate a $U(1)$ invariance under a spacetime dependent change of the
phase instead: \be \Psi \longrightarrow e^{i\phi(x)}\Psi \,,\qquad
\phi=\phi(x)\,?\label{soft}\ee If one does it, the original invariance
of the observables is lost under the new `soft' transformations. In
order to kill the invariance violating terms, one has to introduce a
compensating potential one-form $A$ with values in the Lie algebra of
$U(1)$, which transforms under the soft transformations in a suitable
form. This couples $A$ in a well determined way to the wave function
of the electron, and, if one insists that the $U(1)$-potential $A$ has
its own physical degrees of freedom, then the field strength $F:=d A$
is non-vanishing and the coupled Dirac Lagrangian has to be amended
with a kinetic term quadratic in $F$. In this way one can reconstruct
the whole classical Dirac-Maxwell theory from the naked Dirac equation
together with the postulate of soft phase invariance. Because of
Weyl's original terminology, one still talks about $U(1)$-gauge
invariance -- and not about $U(1)$-phase invariance, what it really is.
Thus the electromagnetic potential is an appendage to the Dirac field
and not related to length recalibration as Weyl originally thought.
\par
\subsection{Yang-Mills and the structure of a gauge theory}
Yang and Mills, in 1954, generalized the Abelian $U(1)$-gauge invariance
to non-Abelian $SU(2)$-gauge invariance, taking the (approximately) {\it
  conserved} isotopic spin {\it current} as their starting point, and,
in 1956, Utiyama set up a formalism for the gauging of any semi-simple
Lie group, including the Lorentz group $SO(1,3)$. The latter group he
considered as essential in GR. We will come back to this topic below.
\par 
In any case, the gauge principle historically originated from GR as a
concept for removing as many action-at-a-distance concept as possible
-- as long as the group under consideration is linked to a conserved
current. 
\begin{figure}
  \epsfbox[-70 -10 500 270]{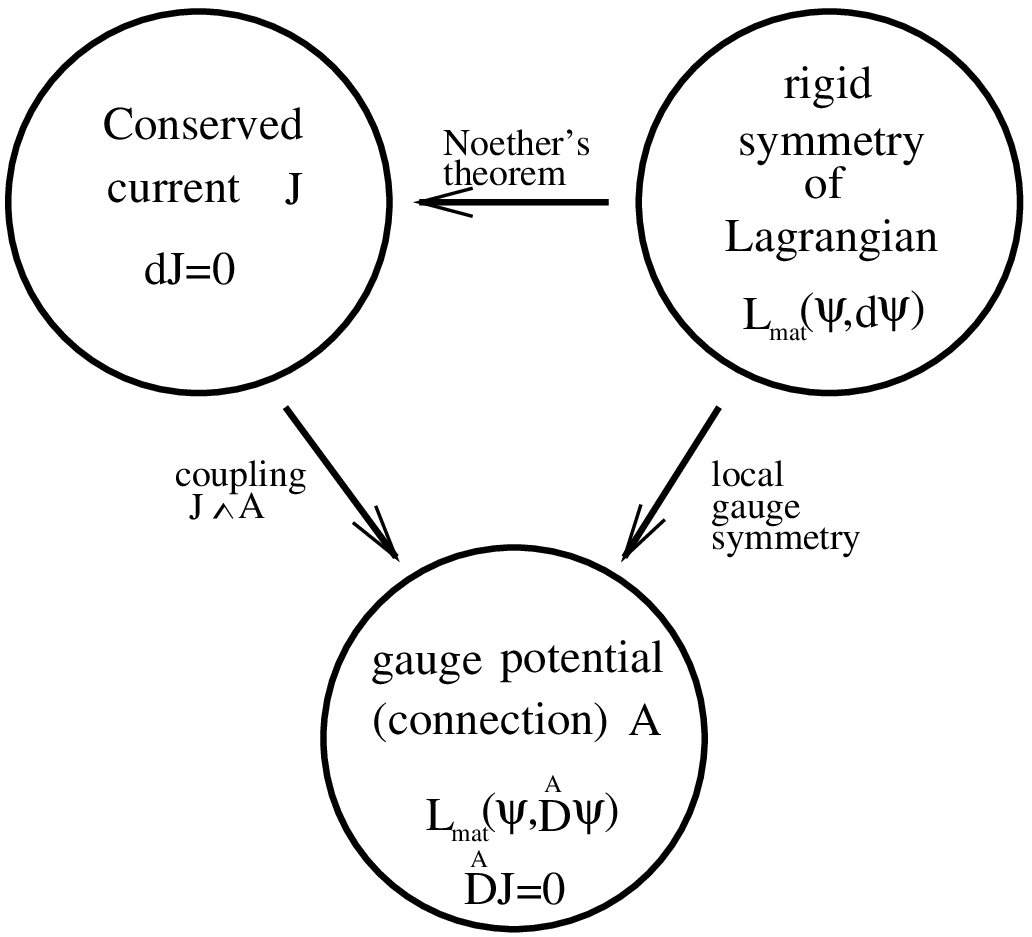} \fcaption{The structure of a gauge
    theory \`a la Yang-Mills is depicted in this diagram, which is
    adapted from Mills\cite{Mills}.  Let us quote some of his
    statements on gauge theories: `The gauge principle, which might
    also be described as a principle of {\em local symmetry}, is a
    statement about the invariance properties of physical laws. It
    requires that {\em every continuous symmetry be a local
      symmetry}...' `The idea at the core of gauge theory...is the
    local symmetry principle: {\em Every continuous symmetry of nature
      is a local symmetry.'} The history of gauge theory has been
    traced back to its beginnings by O'Raifeartaigh\cite{oraifhist}, 
who also gave a compact review of its formalism\cite{ORaif}.}
\label{fig:5}
\end{figure}
This existence of a conserved current of some matter field $\Psi$ 
is absolutely vital for the setting-up of a gauge theory. In 
Fig.\ref{fig:5} we sketched the structure underlying a gauge theory: A 
rigid symmetry of a Lagrangian induces, via Noether's theorem, a 
conserved current $J$, $dJ=0$. It can happen, however, as it did in the 
electromagnetic and the $SU(2)$-case, that a conserved current is 
discovered first and then the symmetry deduced by a kind of a 
reciprocal Noether theorem (which is not strictly valid). 
Generalizing from the gauge approach to the Dirac-Maxwell theory,
we continue with the following gauge procedure:
\par
Extending the rigid symmetry to a soft symmetry amounts to turn the constant
group parameters $\ve$ of the symmetry transformation on the
fields $\Psi$ to functions of spacetime, $\ve\rightarrow\ve(x)$.
This affects the transformation behavior of the matter Lagrangian which 
usually contains derivatives $d\Psi$ of the field $\Psi$: The soft 
symmetry transformations on $d\Psi$ generate terms containing 
derivatives $d\ve(x)$ of the spacetime-dependent group parameters
which spoil the former rigid invariance. In order to counterbalance
these terms, one is forced to introduce a compensating field 
$A=A_i{}^a\tau_a dx^i$ ($a$=Lie-algebra index, $\tau_a$=generators of the
symmetry group) -- nowadays called 
{\it gauge potential} -- into the theory. The one-form $A$ 
turns out to have the mathematical meaning 
of a Lie-algebra valued {\it connection}. It acts on
the components of the fields $\Psi$ with respect to some 
reference frame, indicating that it can be properly represented
as the connection of a frame bundle which is associated to the 
symmetry group.
Thereby it is possible to replace in the matter Lagrangian the 
exterior derivative of the matter field by a gauge-covariant 
exterior derivative,
\be 
d\longrightarrow {\buildrel {A}\over{D}}:=d+A\,,\quad L_{\rm mat}(\Psi, d\Psi) 
\longrightarrow L_{\rm   mat}(\Psi, {\buildrel {A}\over{D}}\Psi)\,.  
\label{mincoupl}
\ee This is called {\it minimal coupling} of the matter field to the
new gauge interaction. The connection $A$ is made to a true dynamical
variable by adding a corresponding kinematic term $V$ to the minimally
coupled matter Lagrangian. This supplementary term has to be gauge
invariant such that the gauge invariance of the action is kept. Gauge
invariance of $V$ is obtained by constructing it from the field
strength $F={\buildrel {A}\over{D}}A$, $V=V(F)$.  Hence the gauge
Lagrangian $V$, as in Maxwell's theory, is assumed to depend only on
$F=dA$, not, however, on its derivatives $d F,\,d\hodge d\,F,\dots$
Therefore the field equation will be of second order in the gauge
potential $A$. In order to make it quasilinear, that is, linear in the
second derivatives of $A$, the gauge Lagrangian must depend on $F$ no
more than quadratically. Accordingly, with the general ansatz
$V=F\wedge H$, where the field momentum or ``excitation'' $H$ is
implicitly defined by $H=-{{\6 V}/{\6 F}}$, the $H$ has to be linear
in $F$ under those circumstances.
\par
By construction, the gauge potential in the Lagrangians couples 
to the conserved current one started with -- and the original 
conservation law, in case of a non-Abelian symmetry, gets 
modified and is only gauge covariantly conserved,
\be
dJ=0\quad\longrightarrow\quad {\buildrel {A}\over{D}}J=0\,,\;\; 
J={{\6 L_{\rm mat}}/{\6 A}}\,. \label{currcoupl}
\ee
The physical 
reason for this modification is that the gauge potential itself 
contributes a piece to the current, that is, the gauge field (in 
the non-Abelian case) is charged. For instance, the Yang-Mills 
gauge potential $B^a$ carries isotopic spin, since the 
$SU(2)$-group is non-Abelian, whereas the electromagnetic potential, 
being $U(1)$-valued and Abelian, is electrically uncharged. 
\par

\subsection{Gravity and the Utiyama-Sciama-Kibble approach}

Let us come back to Utiyama (1956). He gauged the Lorentz group
$SO(1,3)$, inter alia. Using some ad hoc assumptions, like the
postulate of the symmetry of the connection, he was able to recover
GR. This procedure is not completely satisfactory, as is also obvious
from the fact that the conserved current, linked to the Lorentz group,
is the angular momentum current. And this current alone cannot
represent the source of gravity. Accordingly, it was soon pointed out
by Sciama and Kibble (1961) that it is really the {\it Poincar\'e
  group} $R^4\semidirect SO(1,3)$, the semi-direct product of the
translation and the Lorentz group, which underlies gravity. They found
a slight generalization of GR, the so-called Einstein-Cartan theory
(EC), which relates -- in a Einsteinian manner -- the mass-energy of
matter to the curvature and -- in a novel way -- the material {\it
  spin} to the torsion of spacetime. In contrast to the Weyl
connection (\ref{GammaW}), the spacetime in EC is still metric
compatible, i.e.\ governed by a Riemann-Cartan\footnote{The
  terminology is not quite uniform. Borzeskowski and
  Treder\cite{Borz}, in their critical evaluation of different
  gravitational variational principles, call such a geometry a
  Weyl-Cartan gemetry.} (RC) geometry.  Torsion is admitted according to
\begin{equation}
\Gamma^{\rm RC}_{\alpha\beta}=\Gamma^{\{\}}_{\alpha\beta}- {1\over 2}
\left[e_\alpha\rfloor T_\beta-e_\beta\rfloor T_\alpha-(e_\alpha\rfloor
e_\beta\rfloor T_\gamma)\vartheta^\gamma\right]
\,.\label{GammaRC}\end{equation}
Incidentally, approaches to the gauging of the Poincar\'e group on a
{\it fixed} Minkowski background yield effectively similar structures,
see Wiesendanger\cite{Wiesen}.
\par 
In order to fix the notation, let us shortly recapitulate the
structures emerging in spacetime geometry. A four-dimensional
differential manifold has at each point a tangent space, spanned by the
four basis vectors $e_\alpha=e^i{}_\alpha\partial_i$, with
$\partial_i$ as the vectors tangent to the coordinate lines and
$\alpha, \beta=0,1,2,3$, and the cotangent space, spanned by the four
one-forms $\vartheta^\beta=e_j{}^\beta dx^j$, with $dx^j$ as
coordinate one-forms. The two types of bases are dual to each other
$e_\alpha\rfloor\vartheta^\beta=\delta_\alpha^\beta$, where $\rfloor$
denotes the interior product, i.e., in components, $e^k{}_\alpha\,
e_k{}^\beta=\delta_\alpha^\beta$. This is the underlying manifold.  

\par
On top of it we specify a linear connection. For
Einstein\cite{einstein} ``... the essential achievement of general
relativity, namely to overcome `rigid' space (ie the inertial frame),
is {\it only indirectly} connected with the introduction of a
Riemannian metric.  The directly relevant conceptual element is the
`displacement field' ($\Gamma^l_{ik}$), which expresses the
infinitesimal displacement of vectors. It is this which replaces the
parallelism of spatially arbitrarily separated vectors fixed by the
inertial frame (ie the equality of corresponding components) by an
infinitesimal operation.  This makes it possible to construct tensors
by differentiation and hence to dispense with the introduction of
`rigid' space (the inertial frame). In the face of this, it seems to
be of secondary importance in some sense that some particular $\Gamma$
field can be deduced from a Riemannian metric...'' In this vein, we
introduce a linear connection
\be\Gamma_\alpha{}^\beta=\Gamma_{i\alpha}{}^\beta
dx^i\,,\label{conn}\ee with values in the Lie-algebra of the linear
group $GL(4,R)$. These 64 components $\Gamma_{i\alpha}{}^\beta(x)$ of
the `displacement' field enable us, as pointed out in the quotation by
Einstein, to get rid of the rigid spacetime structure of special
relativity (SR).
\par
In order to be able to recover SR in some limit, the primary structure
of a connection of spacetime has to be enriched by the secondary
structure of a metric \be
g=g_{\alpha\beta}\,\vartheta^\alpha\otimes\vartheta^\beta\,,\label{metric}\ee
with its 10 component fields $g_{\alpha\beta}(x)$. At least at the
present stage of our knowledge, this additional postulate of the
existence of a metric seems to lead to the only practicable way to set
up a theory of gravity. In some future time one may be able to
`deduce' the metric from the connection and some extremal property of
the action function -- and some people have tried to develop such type
of models, but without success so far.

\subsection{E.\ Cartan's analysis of general relativity and its consequences}

Besides the gauge theoretical line of development which, with respect
to gravity, culminated in the Sciame-Kibble approach, there was a
second line dominated by E.\ Cartan's (1923) geometrical analysis of
GR. The concept of a linear connection as an independent and primary
structure of spacetime, see (\ref{conn}), developed gradually around
1920 from the work of Hessenberg, Levi-Civita, Weyl, Schouten,
Eddington, and others. In its full generality it can be found in
Cartan's work.  In particular, he introduced the notion of a so-called
torsion -- in holonomic coordinates this is the antisymmetric and
therefore tensorial part of the components of the connection -- and
discussed Weyl's unified field theory from a geometrical point of
view.
\par
For this purpose, let us tentatively call \be
\left(g_{\alpha\beta},\,\vartheta^\alpha,\,\Gamma_\alpha{}^\beta\right)
\label{pot}\ee the {\it potentials} in a gauge approach to gravity and 
\be\left(Q_{\alpha\beta},\,T^\alpha,\,R_\alpha{}^\beta\right)\ee the
corresponding {\it field strengths}. Later, in Sec.\ 6, inter alia, we
will see why this choice of language is appropriate. Here we defined
\ba {\rm nonmetricity}\qquad
Q_{\alpha\beta}&:=&-\stackrel{\Gamma}{D}g_{\alpha\beta}\,,
      \label{nonmetricity}\\
{\rm torsion}\quad\qquad T^\alpha&:=&\stackrel{\Gamma}{D}\vartheta^\alpha=
d\vartheta^\alpha+\Gamma
_\beta{}^\alpha\wedge\vartheta^\beta \,,\label{torsion}\\
{\rm curvature}\qquad R_\alpha{}^\beta &:=&{}^{\prime\prime}
\stackrel{\Gamma}{D}
\Gamma_\alpha{}^\beta{}      
^{\;\prime\prime} =
d\Gamma_\alpha{}^\beta-\Gamma_\alpha{}^\gamma\wedge\Gamma_\gamma{}^\beta
\,.\label{curvature}\ea
Then symbolically we have 
\be \left(Q_{\alpha\beta},\,T^\alpha,\,R_\alpha{}^\beta\right)\;\sim\;
\stackrel{\Gamma}{D}\left(g_{\alpha\beta},\,\vartheta^\alpha,\,
\Gamma_\alpha{}^\beta\right)
\,.\label{field}\ee

\par
By means of the field strengths it is straightforward of how to
classify the spacetime manifolds of the different theories discussed
so far: \ba {\rm GR\; (1915):}\qquad Q_{\alpha\beta}&=&0\,, \qquad
T^\alpha=0 \,, \qquad R_\alpha{}^\beta\neq0\,.\\ {\rm Weyl\;(1918):}
\qquad Q_{\gamma}{}^{\gamma}&\neq&0\,,\qquad T^\alpha=0\,, \qquad
R_\alpha{}^\beta\neq0\,.\\ {\rm EC\;(1923/61):}\qquad
Q_{\alpha\beta}&=&0\,, \qquad T^\alpha\neq0\,, \qquad
R_\alpha{}^\beta\neq0\,.\ea Note that Weyl's theory of 1918 requires
only a nonvanishing {\it trace} of the nonmetricity, the {\it Weyl
  covector} $Q:=Q_\gamma{}^\gamma/4$.  For later use we amend this
table with the Einsteinian teleparallelism (GR$_{||}$), which was
discussed between Einstein and Cartan in considerable detail (see
Debever\cite{Deb}) and with metric-affine gravity\cite{PRs} (MAG),
which presupposes the existence of a connection {\it and} a
(symmetric) metric that are completely independent from each other (as
long as the field equations are not solved): \ba {\rm GR_{||}\;
  (1928):}\qquad Q_{\alpha\beta}&=&0\,, \qquad T^\alpha\neq 0 \,,
\qquad R_\alpha{}^\beta = 0\,.\\ {\rm MAG\;(1976):} \qquad
Q_{\alpha\beta}&\neq&0\,,\qquad T^\alpha\neq 0\,, \qquad
R_\alpha{}^\beta\neq0\,.\ea
\par
Both theories, GR$_{||}$ and MAG, were originally devised as unified
field theories with no sources on the right hand sides of their field
equations. Today, however, we understand them\cite{Cho,PRs} as gauge type
theories with well-defined sources. 

\par
Cartan gave a beautiful geometrical interpretation of the notions of
torsion and curvature. Consider a vector at some point of a manifold,
that is equipped with a connection, and displace it around an
infinitesimal (closed) loop by means of the connection such that the
(flat) tangent space, where the vector `lives' in, rolls without
gliding around the loop. At the end of the journey\cite{PRs} the loop,
mapped into the tangent space, has a small closure failure, i.e.\ a
translational misfit. Moreover, in the case of vanishing nonmetricity
$Q_{\alpha\beta}=0$, the vector underwent a small rotation or -- if no
metric exists -- a small linear transformation. The torsion of the
underlying manifold is a measure for the emerging translation and the
curvature for the rotation (or linear transformation): \ba {\rm
  translation}&\longrightarrow &{\rm torsion}\; T^\alpha \\ {\rm
  rotation\; (lin.\; transf.)}&\longrightarrow &{\rm curvature}
\;R_\alpha{}^\beta\,.\ea Hence, if your friend tells you that he
discovered that torsion is closely related to electromagnetism or to
some other nongravitational field -- and there are many such `friends'
around, as we can tell you as referees -- then you say: `No, torsion
is related to translations, as had been already found by Cartan in
1923.' And translations -- we hope that we don't tell you a secret --
are, via Noether's theorem, related to energy-momentum\footnote{ Not
  long ago, one of the editors of Physical Review D, Lowell S.\ 
  Brown, decreed that all papers with (Cartan's) torsion in the title
  or the abstract are automatically to be rejected. You may guess how
  the authors of such papers reacted. I was always wondering what
  this clever physicist would do, if `energy-momentum' appeared in a
  title. My guess being that he would not even recognize the need for
  becoming active again.}, i.e.\ to the source of gravity, and to
nothing else. We will come back to this discussion in Sec.4.
\par

For the rest of these lectures, unless stated otherwise, we will 
choose the frame $e_\alpha$, and hence also the coframe 
$\vartheta^\beta$, to be {\it orthonormal}, that is, 
\be
g(e_\alpha,e_\beta)\stackrel{*}{=}o_{\alpha\beta}:={\rm diag}(-+++)
\,.\label{ortho}\ee
Then, in a Riemann-Cartan space, we have the convenient antisymmetries 
\be
\Gamma^{RC}_{\alpha\beta}\stackrel{*}{=}-\Gamma^{RC}_{\beta\alpha}
\qquad{\rm and}\qquad 
R^{RC}_{\alpha\beta}\stackrel{*}{=}-R^{RC}_{\beta\alpha}\,.\ee

\section{Einstein's and the gauge approach to gravity}

\subsection{Neutron matter waves in the gravitational field}
\begin{figure}
\vskip 8truecm
\fcaption{The neutron interferometer of the
  COW-experiment\cite{COW,Green}: A neutron beam is split into two
  beams which travel in different gravitational potentials. Eventually
  the two beams are reunited and their relative phase shift is
  measured.}
\label{COWfig1}
\end{figure}

Twenty years ago a new epoch began in gravity: {\bf C}olella-{\bf
  O}verhauser-{\bf W}erner measured by interferometric methods a phase
shift of the wave function of a neutron caused by the gravitational
field of the earth, see Fig.\ref{COWfig1}. The effect could be
predicted by studying the Schr\"odinger equation of the neutron wave
function in an external Newtonian potential -- and this had been
verified by experiment. In this sense nothing really earth-shaking
happened. However, for the first time a gravitational effect had been
measured the numerical value of which depends on the Planck constant
$\hbar$. Quantum mechanics was indispensable in deriving this phase
shift \be \theta_{\rm grav}=\frac{m^2g}{2\pi\hbar^2}\,\lambda\,
A\sin\alpha\,\label{COWphase}\ee ($m$ = mass of the neutron, $\lambda$
its de Broglie wave length, $g$ = gravitational acceleration, $A$ =
area surrounded by the neutron beams, $\alpha$ = angle between the
normal vector of the area $A$ and the vector $\mbold{ g}$).

\begin{figure}
\epsfbox[-30 -10 500 190]{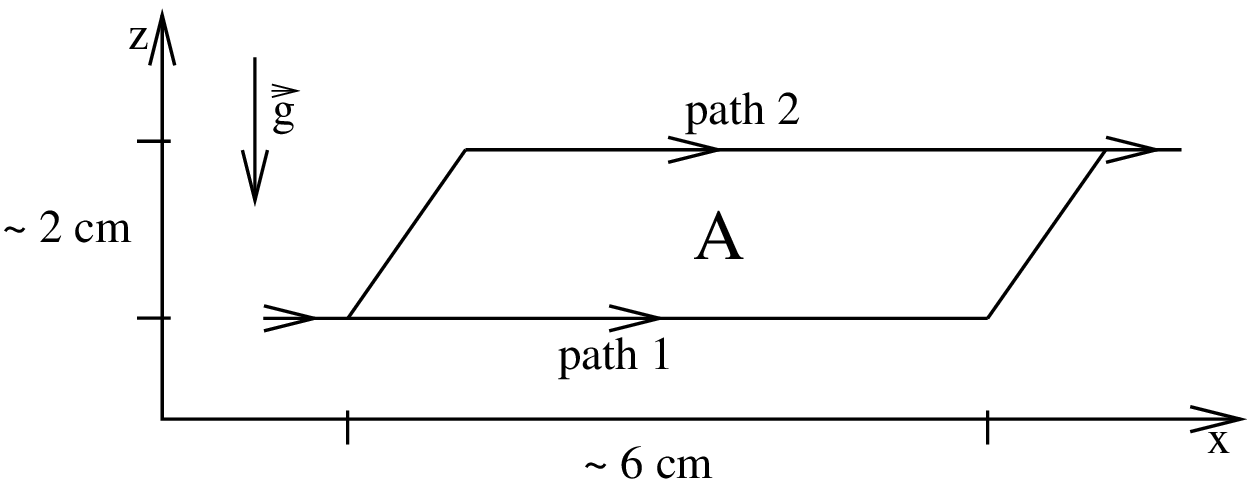}
\fcaption{COW experiment schematically.}
\label{COWfig2}
\end{figure}

It was the availability of nearly perfect single silicon crystals of
about $10\,cm$ length that provided a new tool for X-ray and neutron
interferometry. This had first been demonstrated by Bonse and Hart in
1965 for X-rays. After Bonse (1974) and Rauch, Treimer, and Bonse
(1974), had shown that this device also works for neutrons, Colella,
Overhauser, and Werner (in the following abbreviated by COW) ``...used
a neutron interferometer to observe the quantum-mechanical phase shift
of neutrons caused by their interaction with the Earth's gravitational
field''\cite{COW}. Their experiment is sketched in Fig.\ref{COWfig2}.
\par
They used neutrons cooled to room temperature such that their
resulting mean velocity $v_{n}\simeq 10^{-5}\,c$ is non-relativistic.
Their mass is $m_{n}=1.67\times 10^{-21}\,kg$, and the de Broglie wave
length $\lambda_{n}:=2\pi\hbar/ p\approx 0.2\,nm$. A beam of $1\,cm$
width enters the first `ear' of the interferometer at a Bragg angle in
the range of $20^{\circ}$ to $30^{\circ}$. It is {\it coherently}
scattered by planes of atoms perpendicular to the surface of the
crystal. This Laue scattering gives rise to a transmitted and a
diffracted beam, with opposite Bragg angles. Due to the Borrman
effect, the beam travels through the crystal at first along the planes
and the splitting occurs actually only after it emerges from the ear
again.

\par When the interferometer gets rotated in the gravitational
field of the earth, the upper and lower beams travel at a vertical
distance of about $2\,cm$ and encounter a potential difference
of $\Delta\varphi/E_{\rm kin}=(m_ng\triangle h/(1/2)m_nv^2)\approx
10^{3}/(3\times 10^5)^2\approx 10^{-8}$, which is only a tiny fraction
of the kinetic energy. Nevertheless, this leads to a measurable effect
on the {\it phase} of the neutron's coherent wave which oscillates
about $10\,cm/\lambda_n\approx 10^9$ times during the horizontal
flight.  Although the oscillation rate of the upper beam is
`redshifted' merely by a factor of $10^{-7}$, the upper beam manages
to make $\theta_{\rm grav}\approx 10^9/10^7=100$ oscillations more
than the lower beam.  This phase shift can be observed by the
interference pattern of the recombined beams.

\par In the actual
experiment, side effects have to be taken care of: Gravity produces
distortions in the single crystal. Contributions from this can be
eliminated by comparing X-ray and neutron interference patterns in the
same interferometer. Moreover, the neutron beam itself is bent into a
parabolic path with $4\times 10^{-7}cm$ loss in altitude. This yields,
however, no significant influence on the phase.  

\par In the COW experiment, the single-crystal interferometer is
at rest with respect to the laboratory, whereas the neutrons are
subject to the gravitational potential. In order to compare this with
the effect of acceleration relative to the laboratory frame, {\bf
  B}onse and {\bf W}roblewski\cite{BW} let the interferometer
oscillate horizontally by driving it via a pair of standard
loudspeaker magnets.  Thus these experiments of BW and COW test the
effect of local acceleration and local gravity on matter waves and
prove its {\it equivalence} up to an accuracy of about $4\%$.

\subsection{Accelerated and rotating reference frame}

In order to be able to describe the interferometer in an accelerated
frame, we first have to construct a non-inertial frame of reference.
If we consider only {\it mass points}, then a non-inertial frame in
the Minkowski space of SR is represented by a {\it curvilinear}
coordinate system, as recognized by Einstein\cite{meaning}. Einstein
even uses the names `curvilinear co-ordinate system' and `non-inertial
system' interchangeably.
\par 
According to the standard gauge model of electro-weak and strong
interactions, a neutron is not a fundamental particle, but consists of
{\it one} up and {\it two} down quarks which are kept together via the
virtual exchange of gluons, the vector bosons of quantum
chromodynamics, in a permanent `confinement phase'. For studying the
properties of the neutron in a non-inertial frame and in low-energy
gravity, we may disregard its extension of about $0.7\,fm$, its form
factors, etc. In fact, for our purpose, it is sufficient to treat it
as a Dirac particle which carries spin $1/2$ but is {\it structureless
  otherwise}.
\par
\begin{table}
  \tcaption{Einstein's approach to GR as compared to the gauge
    approach: Used are a mass point $m$ or a Dirac matter field $\Psi$ 
    (referred to a local frame), respectively. IF means inertial frame,
    NIF non-inertial frame. The table refers to special relativity up
    to the second boldface horizontal line. Below, gravity will be switched
    on.  Note that for the Dirac spinor already the force-free motion
    in an {\it inertial} frame {\it does depend} on the mass parameter
    $m$.}
\label{SR}
{\footnotesize 
\bigskip\bigskip
$$ \
\offinterlineskip \tabskip=0pt
\vbox{
\halign
{\strut
\vrule #
\tabskip=0pt plus 100pt
& \hfil # \hfil
& \vrule #
& \hfil # \hfil
& \vrule #
& \hfil  # \hfil
& \vrule #
  \cr
\noalign{\hrule}
&
 && $\matrix{\cr \hbox{\rm \bf Einstein's approach} \cr \cr}$
 && $\matrix{\cr \;\,\hbox{\rm \bf gauge approach} $\,$
   {\rm (}\!\rightarrow{\rm COW)} 
     \cr \cr}$
& \cr
\noalign{\hrule}\noalign{\hrule}\noalign{\hrule}\noalign{\hrule}
&   $\matrix{\cr\;\hbox{\rm elementary object in SR}\cr\cr}$
 && \hbox{\rm mass point m}
 && \hbox{\rm Dirac spinor} $\Psi(x)$
&\cr
\noalign{\hrule}
&    $\matrix{\cr\hbox{\rm inertial frame }\cr\cr}$
 &&  $\matrix{\cr\;\hbox{\rm Cartesian coord.\ system}\ x^i\cr
              ds^2\stareq o_{ij}\,dx^idx^j\cr\cr}$
 &&  $\matrix{\cr\hbox{\rm holonomic orthon.\ frame}\cr
               e_\alpha=\delta_\alpha^i\,\partial_i,\quad e_\alpha\!\cdot
               e_\beta=o_{\alpha\beta}\cr\cr}$
& \cr
\noalign{\hrule}
&    $\matrix{\cr\hbox{\rm force-free}\cr
               \hbox{\rm motion in IF}\cr\cr}$
 &&  $\matrix{\cr{\dot u}{}^i\stareq 0\cr\cr}$
 &&  $\matrix{\cr(i\gamma^i\partial_i-m)\Psi\stareq 0\cr\cr}$
& \cr
\noalign{\hrule}
&    $\matrix{\cr\hbox{\rm non-inertial frame}\cr\cr}$
 &&  $\matrix{\cr\hbox{\rm arbitrary curvilinear}\cr
              \hbox{\rm coord.\ system}\ x^{i'}\cr\cr}$
 &&  $\matrix{\cr\hbox{\rm anholonomic orthon.\ frame}\cr
                e_\alpha=e^i{}_\alpha\,\partial_i\cr
            \hbox{\rm coframe}\ \vartheta^\alpha=e_i{}^\alpha dx^i\cr\cr}$
& \cr
\noalign{\hrule}
&    $\matrix{\cr\hbox{\rm force-free}\cr
              \hbox{motion in NIF}\cr\cr}$
 &&  $\matrix{\cr{\dot u}{}^{i'}+u^{j'}u^{k'}
{\chris{i'}{j'}{k'}}=0\cr
              \cr}$
 &&  $\matrix{\cr\left[i\gamma^\alpha e^i{}_\alpha(\partial_i+\Gamma_i)
     -m\right]\!\Psi=0\cr
           \!   \Gamma_i:={1\over 2}\,\Gamma_i{}^{\beta\gamma}
\rho_{\beta\gamma}
             \quad {\rm Lorentz} 
              \cr\cr}$
& \cr
\noalign{\hrule}
&   $\matrix{\cr\hbox{\rm non-inertial objects}\cr\cr}$
 && $\matrix{\cr{\chris{i'}{j'}{k'}}\cr 40\cr\cr}$
 && $\matrix{\cr\vartheta^\alpha, \ \ \ \ \Gamma^{\alpha\beta}
             =-\Gamma^{\beta\alpha}\cr
                16\ \ \ +\ \ \ \  24\ \ \ \ \ \ \ \cr\cr}$
& \cr
\noalign{\hrule}
&   $\matrix{\cr\hbox{\rm constraints in SR}\cr\cr}$
 && $\matrix{\cr {\tilde R}(\partial\{\},\{ \})=0\cr 20\cr\cr}$
 && $\matrix{\cr T(\partial e,e,\Gamma)\!=\!0,\ R(\partial\Gamma,\Gamma)\!=\!0\cr
                  24\ \ \ \ \ \ \ + \ \ \ \ 36\ \ \cr\cr}$
& \cr
\noalign{\hrule}
&   $\matrix{\cr\hbox{\rm global IF}\cr\cr}$
 && $\matrix{\cr g_{ij}\stareq o_{ij}\,,\ {\chris{i}{j}{k}}\stareq 0\cr\cr}$
 && $\matrix{\cr\left(e_i{}^\alpha,\ \Gamma_i{}^{\alpha\beta}\right)\stareq
    \left(\delta^\alpha_i,0\right)\cr\cr}$
& \cr
\noalign{\hrule}\noalign{\hrule}\noalign{\hrule}\noalign{\hrule}
&    $\matrix{\cr\hbox{\rm switch on gravity}\cr\cr}$
 &&  $\matrix{\cr {\tilde R}\neq 0\cr Riemann\cr\cr}$
 &&  $\matrix{\cr T\neq 0,\ \ R\neq 0\cr Riemann-Cartan\cr\cr}$
& \cr
\noalign{\hrule}
&    $\matrix{\cr\hbox{\rm local IF}\cr
             \hbox{\rm (`Einstein elevator')}\cr
        \cr}$
 &&  $\matrix{\cr g_{ij}\vert_P \stareq o_{ij}\,,\ {\chris{i}{j}{k}}\vert_P
         \stareq  0\cr\cr}$
 &&  $\matrix{\cr(e_i{}^\alpha,\ \Gamma^{\alpha\beta}_i)\vert_P \stareq
              (\delta^\alpha_i,0)\cr\cr}$
& \cr
\noalign{\hrule}
&   $\matrix{\cr\cr\hbox{\rm field equations}\cr\cr\cr}$
 &&  $\matrix{\cr {\tilde R}ic\!-\!{1\over 2}tr({\tilde R}ic)
       \sim mass\cr\cr
         {\bf GR}\cr\cr}$
 &&  $\matrix{\cr\;\, Ric-{1\over 2}tr(Ric)\sim mass\cr
              Tor+2\ tr(Tor)\sim spin\cr
              {\bf EC}\cr\cr}$
& \cr
\noalign{\hrule}
}}$$}
\end{table}

A Dirac particle has to be described by means of a four-component Dirac
{\it spinor}. And this spinor is a half-integer representation of the
(covering group $SL(2,C)$ of the) Lorentz group $SO(1,3)$. Therefore
at any one point of spacetime we need an {\it orthonormal reference
  frame} in order to be able to describe the spinor. Thus, as soon as
matter fields are to be represented in spacetime, the notion of a
reference system has to be generalized from Einstein's curvilinear
coordinate frame $\partial_i$ to an arbitrary, in general anholonomic,
orthonormal frame $e_\alpha$, with $e_\alpha\cdot e_\beta=o_{\a\b}$. 
\par
It is possible, of course, to introduce in the Riemannian spacetime of
GR arbitrary orthonormal frames, too.  However, in the heuristic
process of setting up the {\it fundamental structure} of GR, Einstein
and his followers (for a recent example, see the excellent text of
d'Inverno\cite{dInverno}, Secs.9 and 10) restricted themselves to the
discussion of mass points and holonomic (natural) frames. Matter waves
and arbitrary frames are taboo in this discussion. In Table \ref{SR},
in the middle column, we displayed the Einsteinian method.
Conventionally, {\it after} the Riemannian spacetime has been found
and the dust settled, then electrons and neutron and what not, and
their corresponding wave equations, are allowed to enter the scene.
But before, they are ignored. This goes so far that the
well-documented experiments of COW (1975) and BL (1983) -- in contrast
to the folkloric Galileo experiments from the leaning tower --
seemingly are not even mentioned in d'Inverno\cite{dInverno} (1992).
\par
Prugove{\v{c}}ki\cite{Prugo}, one of the lecturers here in Erice at our
school, in his discussion of the classical equivalence principle,
recognizes the decisive importance of orthonormal frames (see his page
52). However, in the end, within his `quantum general relativity'
framework, the good old Levi-Civita connection is singled out again
(see his page 125). This is perhaps not surprising, since he considers
only zero spin states in this context.
\par
We hope that you are convinced by now that we should introduce
arbitrary orthonormal frames in SR in order to represent non-inertial
reference systems for matter waves -- and that this is important for
the setting up of a gravitational gauge theory\cite{Audretsch,laemmer}. The
introduction of accelerated observers and thus of non-inertial frames
is somewhat standard, even if during the Erice school one of the
lecturers argued that those frames are inadmissible. Take the text of
Misner, Thorne, and Wheeler\cite{MTW}. In their Sec.6, you will find
an appropriate discussion. Together with Ni\cite{HehlNi} and in our
Honnef lectures\cite{Honnef} we tailored it for our needs.
\par
Suppose in SR a non-inertial observer locally measures, by means of
the instruments available to him, a three-acceleration $\mbold{a}$ and
a three-angular velocity $\mbold{\omega}$. If the {\it laboratory
  coordinates} of the observer are denoted by $x^{\overline{i}}$, with
$\overline{\mbold{x}}$ as the corresponding three-radius vector, then
the non-inertial frame can be written in the succinct
form\cite{HehlNi,Honnef} \ba e_{\hat{0}}&=&\frac{1}{1+ \mbold{a}
\cdot\overline{\mbold{x}}/c^2}\,
\left[\partial_{\,\overline{0}}-\left(\frac{\mbold{\omega}}{c}\times
\overline{\mbold{x}}\right)^{\overline{B}}
\partial_{\,\overline{B}}\right],\nonumber\\ 
e_{ A}&=&\partial_{\,\overline{A}} \,.\label{labframe}\ea Here `naked'
capital Latin letters, $A,\dots={\hat 1},{\hat 2},{\hat 3}$, denote
spatial anholonomic components.  For completeness we also display the
coframe, that is, the one-form basis, which one finds by inverting the
frame (\ref{labframe}): \ba \vartheta^{\hat 0}&=&\left(1+\frac{\mbold{a}
\cdot\overline{\mbold{x}}}{c^2}\right)\,dx^{\overline{0}}
=N\,dx^{\overline{0}}\,,\nonumber\\ \vartheta^{A}&=&dx^{\overline{A}}
+\left(\frac{{\mbold{\omega}}}{c} \times\overline{\mbold{x}}
\right)^{\overline{A}}\,dx^{\overline{0}} =dx^{\overline{A}}
+N^{\overline{A}}\,dx^{\overline{0}}\,.
\label{labcoframe}\ea 
In the $(3+1)$-decomposition of spacetime, $N$ and
$N^{\overline{A}}$ are known as {\it lapse function} and {\it shift vector},
respectively. 
\par
Starting with the coframe, we can read off the connection coefficients
(since torsion vanishes in SR) by using Cartan's first structure
equation
\hbox{$d\vartheta^\alpha=-\Gamma_\beta{}^\alpha\wedge\vartheta^\beta$}
with
$\Gamma_\beta{}^\alpha=\Gamma_{{\bai}\,\beta}{}^\alpha\,dx^{\bai}$.
We find: \ba \Gamma_{\baz\,{\hat 0}A}&=&-\Gamma_{\baz A{\hat 0}}
={a_A\over c^2}\,,\nonumber\\ \Gamma_{\baz AB}&=&-\Gamma_{\baz BA}
=\epsilon_{ABC}\,{\omega^C\over c}\,.\label{labconn1}\ea It is
important to note that the first index is holonomic, whereas the
second and the third indices are anholonomic. If we transform the
first index, by means of the frame coefficients $e^{\bai}{}_\alpha$,
with $e_\a=e^{\overline{i}}{}_\a\,\partial_{\,\overline{i}}$, into an
anholonomic one, then we find the totally anholonomic connection
coefficients as follows: \ba \Gamma_{{\hat 0}{\hat
    0}A}&=&-\Gamma_{{\hat 0}A{\hat 0}} = {a_A\over
  c^2}\,/\left(1+\mbold{a}\cdot\overline{\mbold{x}}/{c^2} \right)
\,,\nonumber \\ \Gamma_{{\hat 0}AB}&=&-\Gamma_{{\hat 0}BA}
=\epsilon_{ABC}\, {\omega^C\over c}\,/\left(1+\mbold{a}
\cdot\overline{\mbold{x}}/{c^2} \right) \,.\label{labconn2} \ea
These connection coefficients (\ref{labconn2}) will enter the Dirac
equation referred to a non-inertial frame.
\par
In order to assure ourselves that we didn't make mistakes in computing
the `non-inertial' connection (\ref{labconn1},\ref{labconn2}) by hand, we
used for checking its correctness the EXCALC package on exterior
differential forms of the computer algebra system REDUCE, see Puntigam
et al.\cite{Puntigam} and the literature given there.

\subsection{Dirac matter waves in a non-inertial frame of reference}

The phase shift (\ref{COWphase}) can be derived from the Schr\"odinger
equation with a Hamilton operator for a point particle in an external
Newton potential. For setting up a gravitational theory, however, one
better starts more generally in the special relativistic domain. Thus
we have to begin with the Dirac equation in an external gravitational
field or, if we expect the equivalence principle to be valid, with the
Dirac equation in an accelerated and rotating, that is, in a
non-inertial frame of reference.
\par
Take the Minkowski spacetime of SR. Specify Cartesian coordinates.
Then the field equation for a massive fermion of spin $1/2$ is
represented by the Dirac equation \be
i\hbar\gamma^i\,\partial_i\psi\stareq mc\psi\,,\label{Dirac1}\ee where the
Dirac matrices $\gamma^i$ fulfill the relation
\be\gamma^i\gamma^j+\gamma^j\gamma^i=2\,o^{ij}\,.\ee For the
conventions and the representation of the $\gamma$'s, we essentially
follow Bjorken-Drell\cite{BjD}.
\par

Now we straightforwardly transform this equation from an inertial to
an accelerated and rotating frame. By analogy with the equation of
motion in an arbitrary frame as well as from gauge theory, we can
infer the result of this transformation: In the non-inertial frame,
the partial derivative in the Dirac equation is simply replaced by the
{\it covariant derivative} \be\partial_i\quad\Rightarrow\quad
D_\alpha:=\partial_\alpha+{i\over 4}\sigma^{\beta\gamma}\,
\Gamma_{\alpha\beta\gamma}\,,\qquad
\partial_\alpha:=e^i{}_\alpha\partial_i\equiv e_{\alpha}\,,\ee where
$\Gamma_{\alpha\beta\gamma}$ are the anholonomic components of the
connection, see (\ref{labconn2}), and $x^{i}$ the Cartesian
coordinates of the lab system (which we called $x^{\bai}$ previously;
we drop the bar for convenience).  The anholonomic Dirac matrices are
defined by
\be\gamma^\alpha:=e_i{}^\alpha\,\gamma^i\quad\Rightarrow\quad
\gamma^\alpha\gamma^\beta+\gamma^\beta\gamma^\alpha=2\,o^{\alpha\beta}\,.
\ee The six matrices $\sigma^{\beta\gamma}$ are the infinitesimal
generators of the Lorentz group and fulfill the commutation relation
\be[\gamma^\alpha,\sigma^{\beta\gamma}]=2i(o^{\alpha\beta}\gamma^{\gamma}
-o^{\alpha\gamma}\gamma^{\beta})\,.\ee For Dirac spinors, the Lorentz
generators can be represented by
\be\sigma^{\beta\gamma}:=(i/2)(\gamma^\beta\gamma^\gamma
-\gamma^\gamma\gamma^\beta)\,,\label{sigma}\ee furthermore,
\be\mbold{\alpha} :=\gamma^{\hat 0}\mbold{\gamma}\qquad{\rm
  with}\qquad\mbold{\gamma}=\{\gamma^\Xi\}\,.\ee Then, the Dirac equation,
formulated in the orthonormal frame of the accelerated and rotating
observer, reads \be i\hbar\gamma^\alpha
D_\alpha\psi=mc\psi\,.\label{Dirac2}\ee Although there appears now a
`minimal coupling' to the connection, which is caused by the change of
frame, there is no new physical concept involved in this equation.
Only for the measuring devices in the non-inertial frame we have to
assume hypotheses similar to the clock hypothesis. This proviso can
always be met by a suitable construction and selection of the devices.
Since we are still in SR, torsion and curvature of spacetime both
remain zero.  Thus (\ref{Dirac2}) is just a reformulation of the
`Cartesian' Dirac equation (\ref{Dirac1}).
\par
The rewriting in terms of the covariant derivative provides us with a
rather elegant way of explicitly calculating the Dirac equation in the
non-inertial frame of an accelerated, rotating observer: Using the
anholonomic connection components of (\ref{labconn2}) as well as
$\mbold{\alpha}=-i\{\sigma^{\hat 0\Xi}\}$, we find for the covariant
derivative: \ba D_{\hat 0}&=&{1\over 1+\mbold{ a}\cdot\mbold{ x}/c^2}
\left(\partial_0+{1\over 2c^2}\,\mbold{ a}\cdot\mbold{\alpha} -{i\over
    c\hbar}\,\mbold{\omega}\cdot\mbold{ J}\right)\,,\nonumber\\ 
D_{\,\Xi}&=&\partial_{\,\Xi}\,.\ea The total three-angular momentum
operator \be\mbold{ J}:=\mbold{ L}+\mbold{ S}=\mbold{
    x}\times{\hbar\over i}\, {\partial\over\partial\mbold{ x}}+{1\over
        2}\,\hbar\,\mbold{\sigma} =\mbold{ x}\times\mbold{ p}+{1\over
              2}\,\hbar\,\mbold{\sigma}\,\ee is composed, in the
              canonical manner, from the orbital piece $\mbold{ L}$ and
                the spin piece $\mbold{ S}$.
\par
The physical effects in our lab frame can be most easily understood by
going over to the Hamiltonian. After multiplying the Dirac equation by
$\beta:=\gamma^{\hat 0}$ and $c(1+\mbold{ a}\cdot\mbold{ x}/c^2)$, we get \be
i\hbar{\partial\psi\over\partial t}=H\psi\quad\hbox{\rm with } H=\beta
mc^2+{\cal O}+{\cal E}\,.\ee After substituting the covariant
derivatives, the operators ${\cal O}$ and ${\cal E}$, which are odd
and even with respect to $\beta$, read, respectively\cite{HehlNi}:
\be{\cal O}:=c\mbold{\alpha}\cdot\mbold{ p}+{1\over 2c} \left\{(\mbold{
  a}\cdot\mbold{ x})(\mbold{ p}\cdot\mbold{\alpha}) +(\mbold{ p}\cdot
\mbold{\alpha})(\mbold{
  a}\cdot\mbold{ x})\right\}\,,\ee \be{\cal E}:=\beta m(\mbold{
a}\cdot\mbold{ x})-\mbold{\omega}\cdot \left(\mbold{ L}+\mbold{ S}\right)\,.\ee
Up to now, these are mathematically {\it exact} results.  For later
purposes we introduce ${\cal O}_1=c\mbold{\alpha}\cdot\mbold{ p}$ and ${\cal
  O}_2={\cal O}-{\cal O}_1$.

\begin{table}[h]
\tcaption{Inertial effects for a massive fermion of spin $1/2$ in 
non-relativistic approximation.}
\bigskip
\label{effects}
\begin{center}
\small
\begin{tabular}{||c|c||}\hline\hline
&\\
{$\beta\,m\,(\mbold{ a}\cdot\mbold{ x})$} &{Redshift\  (Bonse-Wroblewski 
$\rightarrow$ COW)}\\
&\\
{$-\mbold{\omega}\cdot\mbold{ L}$}
 &{Sagnac type effect\  (Heer-Werner et al.)}\\
&\\
$-\mbold{\omega}\cdot\mbold{ S}$ & Spin-rotation effect\  (Mashhoon)\\
&\\
$\beta\,\mbold{ p}\cdot(\mbold{ a}\cdot\mbold{ x})\ \mbold{ p}\ /\,(2mc^2)$ & 
Redshift effect of kinetic energy \\
&\\
$\beta\,\hbar\,\mbold{\sigma}\cdot(\mbold{ a}\times\mbold{ p})\ /\,(4mc^2)$ &
New inertial spin-orbit coupling \\
&\\
\hline\hline
\end{tabular} 
\end{center}
\end{table}

\par 
Similarly as in quantum electrodynamics\cite{BjD}, a non-relativistic
approximation can be obtained by applying successive Foldy-Wouthuysen
transformations.  After three such steps we find\cite{HehlNi}, up to
the order of $c^{-2}$, \ba H' &=&\beta mc^2+{\beta\over
  2m}\mbold{p}^2-{\beta\over 8m^3c^2}\mbold{ p}^4 +\beta
m(\mbold{a}\cdot\mbold{x})-\mbold{\omega}\cdot(\mbold{L}+\mbold{S})\nonumber\\ 
&+&{\beta\over 2m} \mbold{p}\cdot{\mbold{a}\cdot\mbold{ x}\over
  c^2}\,\mbold{p} +{\beta\hbar\over 4mc^2}\mbold{\sigma}\cdot
\mbold{a}\times\mbold{p} +O({1\over c^3})\,.\ea The different
 non-inertial effects of a fermion are displayed in Table
    \ref{effects}. Besides the rest mass and the usual kinetic term,
    we obtain terms which account for the {\it redshift effect} due to
    acceleration, thereby verifying the BW and, if the acceleration
    $\mbold{a}$ is substituted by the gravitational acceleration
    $\mbold{g}$, the COW experiments.  Moreover the {\it `Sagnac
      type' effect} occurs in the same manner as in the
    non-relativistic Schr\"odinger equation, and a spin-rotation
    effect is found which, for the neutron interferometer, has first
    been proposed by Mashhoon\cite{Mashhoon}. This term could not have
    been obtained by using the Schr\"odinger equation.
\par
Thus we demonstrated that the Dirac equation in a Minkwoski
spacetime\footnote{These considerations can be generalized to a
  Riemannian spacetime, see Huang\cite{Huang} and the literature
  quoted there.}, referred to a non-inertial frame, yields the BW phase
shift and, if the equivalence principle is assumed, also that of the
COW experiment.  The claim of our colleague in Erice, that in SR such
non-inertial frames are illegitimate, is thus disproved. Rather we have
shown the usefulness of non-inertial frames $\vartheta^\a$, see
(\ref{labcoframe}), with $d\vartheta^\a\neq 0$.

\subsection{`Deriving' a theory of gravity: Einstein's method as opposed 
to the gauge procedure}

Now we turn back to Table \ref{SR}: Einstein's approach, in the middle
column, is compared with the gauge approach represented in the right
column. The basic idea is that a mechanical system is considered in an
inertial and then in a non-inertial frame. In Einstein's approach the
object under investigation is a point particle and the non-inertial
frame a {\it natural} (or coordinate) coframe $dx^i$, in the gauge
approach the object is a Dirac wave function and the non-inertial
frame an {\it arbitrary} (or anholonomic) coframe $\vartheta^\a$, with
$d\vartheta^\a\neq 0$. Einstein's `trick' was to search for the fields
emerging in the non-inertial frame and to identify them as
pseudo-gravitational. `Pseudo', since the spacetime is a flat and
uncontorted Minkowski space without gravitational fields present.
This implies that the pseudo-gravitational fields have to obey some
constraints, otherwise it wouldn't be guaranteed that the Minkowski
background prevails.
\par
The gauge approach had been developed by Utiyama et al. long before BW
and COW set up their experiments. These experiments confirmed the
appropriateness of the earlier theoretical development. Clearly, in
some WKB-limit, compare the Foldy-Wouthuysen technique in Sec.\ 3.3,
the Einsteinian approach can be recovered as a limiting case of the
more general gauge approach.
\par
In the Einstein case the pseudo-gravitational field is represented by
the Christoffel symbol (of the second kind), in the gauge case by the
coframe and the Lorentz-connection. The transition from SR to a
spacetime with `real' gravity consists in the {\it relaxation} of the
constraints formulated earlier. We are finding a Riemannian and a
Riemann-Cartan space\footnote{Recently Hammond\cite{ham} gave a very
  pronounced and interesting plea in favor of the existence of torsion
  in nature. Only his attempts to derive torsion from some new
  potential, we don't find convincing, since it looks ad hoc to us.
  Moreover, the coframe is some sort of potential already, see
  (\ref{torsion}) and Table 5 -- and why should we multiply the number
  of potentials beyond necessity?}, respectively.
\par
But beware! Locally we have to recover the Minkowskian behavior of
spacetime: In the Einsteinian free-fall elevator gravity does not show
up. In the Riemann space this is technically achieved by means of
Riemannian normal coordinates, in the case of the Riemann-Cartan space
there exist {\it normal frames} which realize an `anholonomic'
elevator, as has been first pointed out by von der Heyde\cite{vdH1}
and developed by one of us\cite{Erice}, by Modanese \& Toller\cite{Modanese},
Iliev\cite{Iliev}, and Hartley\cite{hart}.  In Hartley's
formulation\cite{hart}, we have the following proposition:
\bigskip
\par
{\it Let M be a manifold with metric g and (arbitrary)
  metric--compatible connection $\nabla $ (Riemann--Cartan space). For
  any single point $P\in M$, there exist coordinates $\lbrace
  x^{i}\rbrace $ and an orthonormal frame $\lbrace e_{\alpha}\rbrace $
  in a neighborhood of $P$ such that \be\left\{\begin{array}{rcl}
      e_{\alpha}&=&\delta _{\alpha}^{i}\,\partial _{x^{i}}\\ \Gamma
      _{\alpha}{}^{\beta}&=&0
         \end{array}
       \right\}\quad {\rm at}\,\,\, P\,, \label{propo}\ee where
       $\Gamma_{\alpha}{}^{\beta}$ are the connection 1--forms
referred to the frame $\lbrace e_{\alpha}\rbrace $.}\bigskip

Not too many people know of this theorem, even though it had been
proposed by von der Heyde already twenty years ago. Usually, if we talk
to relativists about the normal frames in a Riemann-Cartan space, they
state that those cannot exist, since torsion, as a tensor, cannot be
transformed to zero. In this context it is tacitly assumed that the
starting point are Riemannian normal coordinates and the torsion is
`superimposed'. However, since only a {\it natural} frame is attached
to Riemannian normal coordinates, one is too restrictive in the
discussion right from the beginning. And, of course, the curvature is
also of tensorial nature -- and still Riemannian normal coordinates do
exist.
\par
In Table \ref{SR} there are two subtleties involved which we want to
mention:
\par
In the gauge column we formulated two constraints in SR, namely the
vanishing of torsion and that of curvature. One could well wonder
whether we have to relax both constraints at the same time. We could
allow for torsion only, thereby ending up with a spacetime carrying a
teleparallelism, or, alternatively, we could only admit the emergence
of curvature, thereby recovering the Riemannian spacetime of GR. It is
perhaps surprising that also in the teleparallelism case, see Secs.4
and 5 below, by a suitable choice of a torsion square Lagrangian, one
can arrive at a theory which is equivalent to GR.  Nevertheless, there
are good reasons for relaxing both constraints: Firstly, by mimicking
the Einsteinian procedure of the middle column of Table 1, we cannot
see any reason why we should lift only one constraint; secondly, by
admitting a Riemann-Cartan space, we can still `trivialize' the
gravitational potentials at a point $P$ and its neighborhood, as found
in proposition (\ref{propo}), in spite of the presence of torsion
and curvature. This implies that a Riemann-Cartan space, if described
in terms of suitable coframes, looks locally Minkowskian.
Eqs.(\ref{propo}) supply the strongest reason for taking the
four-dimensional Riemann-Cartan space seriously as a model for
spacetime.
\par
A second subtlety is related to the fact that the minimally coupled
Dirac equation (\ref{Dirac2}), in a Riemann-Cartan space, slightly
differs from the Euler-Lagrange equation of the minimally coupled
Dirac {\it Lagrangian}. However, this shouldn't cause headaches: The
gauge theoretical set up, see Fig.\ \ref{fig:5}, is so closely linked
to the Lagrangian formalism -- not to speak of the fundamental
importance of the Feynman path integral or of Schwinger's variational
principle -- that the {\it minimal coupling} procedure should be
implemented on the {\it Lagrangian level}.
\par
Eventually concentrating our attention to the last row of Table
\ref{SR}, we read off the field equations of GR and of the EC-theory:
They result from the simplest conceivable Lagrangian, namely from the
curvature scalar of the Riemann or the Riemann-Cartan spacetime
respectively. The EC-theory differs from GR in a very weak
spin-contact interaction which is unmeasurable at the present time. In
this sense, the EC-theory is a {\it viable} theory of gravitation.

\section{Conserved momentum current, the heuristics of the 
translation gauge}

\subsection{Motivation}
We have in mind to derive gravity from a symmetry principle. But what
is the right symmetry to derive gravity from? As already indicated in
Sec.2, we think that gravity stems from translation symmetry. Our
motivation is the following:
\par
We start from SR. The invariance of the action $W=\int
L_{\rm mat}(\Psi, d\Psi)$ of an isolated material system under rigid
spacetime translations yields, by the application of the Noether
theorem, a {\it conserved energy-momentum current} three-form $\TT_j$
via \be \TT_j:={{\d L_{\rm mat}}\over{\d
    dx^j}}=\frac{1}{3!}\,\TT_{klmj}\,dx^k\wedge dx^l\wedge dx^m\,,\qquad
\qquad d\,\TT_j=0\,. \label{tj} \ee (One obtains the ``usual'' energy-momentum
tensor $T_{ij}$ from $\TT_j$ by means of
$T_{ij}=\epsilon_i{}^{klm}\TT_{klmj}$.) The corresponding charge
$M:=\int d^3x \TT_0$ is conserved in time. In other words: Rigid
translational invariance is attributed to the classical
field-theoretical equivalent to mass(-energy density), which is the
source of Newton-Einstein gravity.
\par
The analogy to electrodynamics guides us of how to actually generate
the gravitational interaction. In electrodynamics one finds from rigid
$U(1)$-invariance of an action $W=\int L_{\rm mat}(\Psi, d\Psi)$ a
conserved electromagnetic current with corresponding electric charge
$Q$.  As we discussed in length in Sec.2, it is possible to generate
the electromagnetic interaction by gauging the rigid $U(1)$-symmetry.
Following this example, we expect the gravitational interaction to
emerge from gauging the rigid translational symmetry. To quote
Feynman\cite{Feyn}: ``...gravity is that field which corresponds to a
gauge invariance with respect to displacement transformations.'' But
before we gauge the translation group we should think about...

\subsection{Active and passive translations}
Whenever we describe quantities with respect to some reference system
(a certain basis, a certain coordinate system) we do encounter the 
possibility of active or passive transformations. Actively transforming a 
quantity or passively transforming its reference system will usually change
the labels of the quantity which are related to the reference system. 
How can we decide whether an active, a passive, or even a mixture of both 
kinds of transformation took place when those labels of a quantity changed?
Is it necessary to distinguish between active and passive transformations?
For the case of active and passive translations, we will answer this question
to the extent that we show the mathematical equivalence of actively or
passively translating a matter field $\Psi$ in Minkowski spacetime. 
{}From this we will later proceed in order to gauge the translation  
group.
\par
The field $\Psi$ is thought to be some $p$-form, possibly Lie-algebra
valued. We discard any internal structure of $\Psi$ since we are only
interested in its spacetime properties. For the Minkowski spacetime we
assume for convenience a gauge in which the connection components are
identically zero, $\G\stackrel{*}{=} 0$.  This can be achieved by
choosing pseudo-Cartesian coordinates $x^i$ and using the
corresponding holonomic (co-)frame.  The total variation $\d_t\Psi$ of
$\Psi$ under translations is the sum of both active and passive
variations: \be \d_t\Psi=\d_a\Psi+\d_p\Psi\,. \label{totalvar} \ee An
active translation of $\Psi$ is generated by the flow of a vector
field $\xi=\xi^i\6_i$, that is, $\Psi\rightarrow \xi^*\Psi$. In this
case the active variation is given by the Lie-derivative $l_\xi$, \be
\d_a\Psi=l_\xi\Psi\,,\qquad
l_\xi=d(\xi\rfloor)+(\xi\rfloor)d\,.\label{actvar} \ee Despite
possible exterior indices of $\Psi$, we don't have to use a `covariant
Lie derivative' since $\Gamma\stackrel{*}{=} 0$. We will be
interested only in infinitesimal variations. Then the Lie derivative
describes the difference between the actively translated
$\xi^*\Psi(x^i)$ and $\Psi(x^i+\xi^i)$.  
\par
Passive translations are
generated by coordinate transformations. To find the passive
transformation which corresponds to a given active transformation, we
have to know the explicit form of the operator $\d_p$ and the actual
coordinate transformation which corresponds to a specific
$\d_a=l_\xi$.  For the actual coordinate transformation, we look at
the total variation of the coordinate functions $x^i$, \be \d_t
x^i=\d_a x^i+\d_p x^i=l_\xi x^i+\d_p x^i=\xi^i+\d_p x^i\,.\label{xvar}
\ee The coordinate transformation $\d x^i=\d_p x^i=-\xi^i$ makes $\d_t
x^i$ to vanish. Thus the passive translation $\d_p x=\xi$ is equivalent
to $\d_a x=l_\xi x$ (what we intuitively expected without any
calculation).  Having this candidate for the appropriate coordinate
transformation, we look for an operator $\d_p$ such that
$\d_t\Psi=l_\xi\Psi +(\d_p\Psi)(-\xi) =0$. We set \be
(\d_p\Psi)(-\xi)=\Psi(x^i-\xi^i)-\Psi(x^i)\,,\label{passvar1} \ee
i.e., we take the value of $\Psi$ at a point in the coordinate system
$x^i-\xi^i$ and subtract the value of $\Psi$ at the same point while
using the coordinate system $x^i$. Writing \be
(\d_p\Psi)(-\xi)=\Psi_{i_1...i_p}(x^i-\xi^i)
d(x^{i_1}-\xi^{i_i})\wedge...\wedge d(x^{i_p}-\xi^{i_p})\,, \ee Taylor
expanding $\Psi_{i_1...i_p}(x^i-\xi^i)$, and keeping terms up to order
$\xi$, we can show that \be (\d_p\Psi)(-\xi)=-l_\xi\Psi\,.
\label{passvar2} \ee Thereby we just rederived the definition of the
Lie derivative from a passive point of view, and the coordinate
transformation $\d x=\xi$ becomes clearly equivalent to an active
translation generated by $l_\xi$.

\subsection{Heuristic scheme of translational gauging}

In order to gauge the translation group we will follow the general
gauge scheme which we set up in Sec.2. Therefore we will begin with a
rigidly translation invariant action $W=\int L_{\rm mat}(\Psi,
d\Psi)$. We expect to have to introduce four gauge potentials $A^\a$
in order to extend from rigid to soft translation invariance of $W$.
Each $A^\a$ will compensate one of the four independent soft
translations $l_{\xi_\a}$.
\par
We note that soft 
translations do not commute any longer: Expanding the flow-generating vectors
$\xi_\a$ according to $\xi_\a=e^i{}_\a\,\6_i$, with spacetime dependent 
functions $e^i{}_\a=e^i{}_\a(x)$, we can use the formula
\be
[l_{\xi_\a},l_{\xi_\b}]\,=\,l_{[{\xi_\a},\xi_\b]} \label{liecomm}
\ee
to show that (the equality ${\buildrel {*}\over{=}}$ requires the 
gauge $\G\stackrel{*}{=} 0$)
\be
[l_{\xi_\a},l_{\xi_\b}]\,=\,-e^j{}_{[\a}\, e^i{}_{\b]}\,
e_i{}^\g{}_{,\,j}\;\xi_\g
\,\stackrel{*}{=}\,-T_{\a\b}{}^\g\,\xi_\g\,.  \label{softcomm}
\ee
The $T_{\a\b}{}^\g$ are the components of the torsion two-form
$T^\g=\frac{1}{2}\,T_{\a\b}{}^\g\vt^\a\wedge\vt^\b$ expanded in the
anholonomic coframe $\vt^\a=e_i{}^\a dx^i$. This softening of the Lie
algebra of the translation group should be seen in correspondence to
the structure \be {\rm commutator}\; {\rm of}\; {\rm soft}\; {\rm
  symmetry}\; {\rm transformations}\,= \,{\rm field}\; {\rm
  strength}\,,\label{commstructure} \ee which is well known from
Yang-Mills theory. Therefore it is tempting to view the torsion tensor
as the translational field strength.
\par
The torsion tensor represents a translational misfit, as already
indicated in Sec.2.4. A more detailed discussion is given by de
Sabbata and Sivaram\cite{Venzo} who also collected numerous other
facts and results on torsion. Torsion measures the noncommutativity of
displacements of points in analogy to the curvature tensor which
measures the noncommutativity of displacements of vectors.
\begin{figure}
  \epsfbox[-70 -10 500 270]{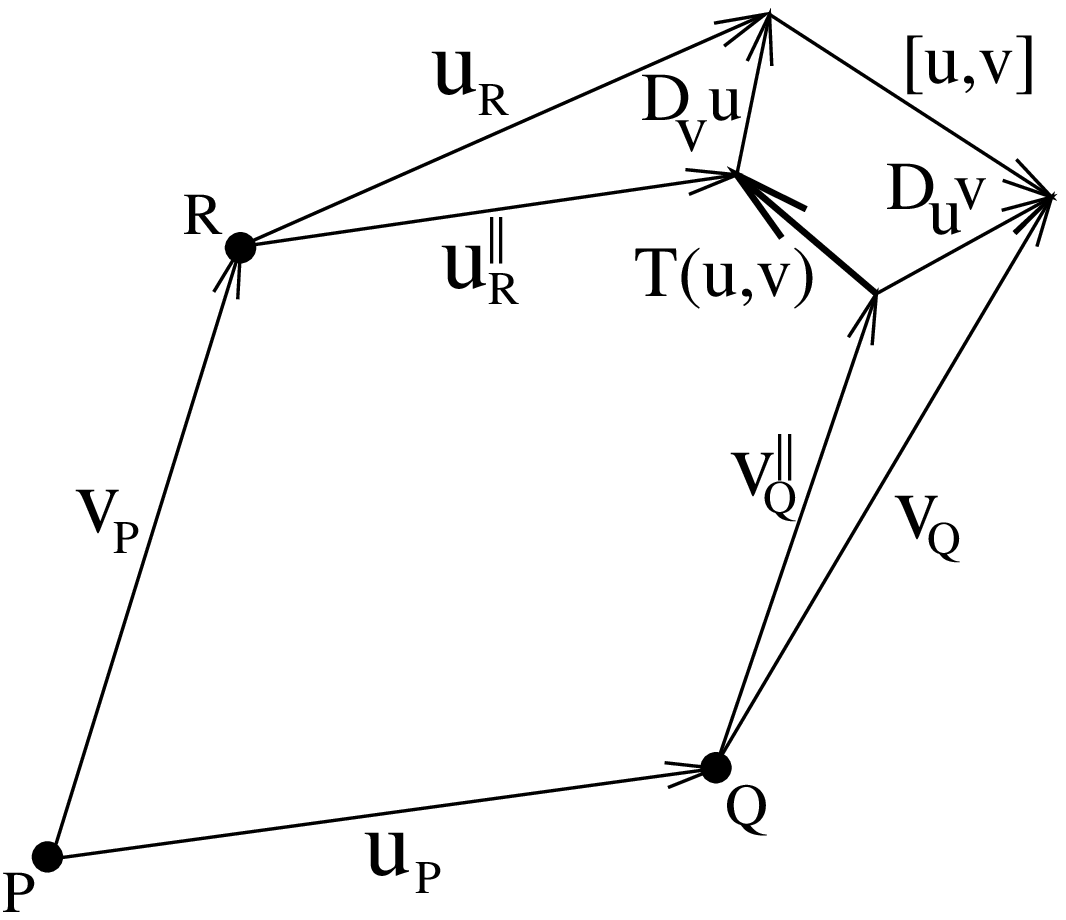} \fcaption{On the geometrical
    interpretation of torsion: It represents a closure failure of
    infinitesimal displacements.}
\label{torsionfig}
\end{figure}
This is explained in Fig.\ \ref{torsionfig}: Let $P$ be a point of the
(spacetime) manifold $M$ and $v_P$, $u_P$ two linearly independent
vectors of $T_PM$. We regard $v_P$, $u_P$ as infinitesimally small.
Then they define two points, $R$ and $Q$, on $M$.  (We can make this
mathematically precise by defining $R$, $Q$ via the exponential map
$\exp_P:T_PM\rightarrow M,\; \exp_P(v)=R,\; \exp_P(u)=Q$.)  In other
words: The vector $v_P$ displaces $P$ infinitesimally to $R$, the
vector $u_P$ displaces $P$ infinitesimally to $Q$. The prescription to
perform two successive displacements $v$, $u$ is to first displace $P$
by means of $v_P$ to $R$ and second parallel transport $u_P$ to
$u_R^{||}$ and displace $R$ by means of $u_R^{||}$. Fig.\ 
\ref{torsionfig} shows that the commutator of two successive
displacements won't vanish in general -- the gap between the two
resulting points is, by definition, a measure for the torsion. It is
also shown in Fig.\ref{torsionfig} that this definition matches the
usual textbook definition \be
T(u,v):=D_uv-D_vu-[u,v]\,.\label{textbookdef} \ee To make contact with
our present notation, we evaluate the components of $T$ with respect
to an arbitrary frame $e_\alpha$: \bea T(e_\b,e_\g) &=&
D_{e_\b}e_\g-D_{e_\g}e_\b -[e_\b,e_\g] \nonumber\\ &=&\G_{\b\g}{}^\a
e_\a-\G_{\g\b}{}^\a e_\a+C_{\b\g}{}^\a e_\a\\ &=&T_{\b\g}{}^\a e_\a
\,.\nonumber \label{torscomp1} \eea Hence \be
T_{\b\g}{}^\a=C_{\b\g}{}^\a+\G_{\b\g}{}^\a-\G_{\g\b}{}^\a\,,\label{torscomp2}
\ee 
or, by using the object of anholonomity \be C^\a:=d\vt^\a={1\over
  2}C_{\b\g}{}^\a\,\vt^\b\wedge\vt^\g\,,\ee we find:\be
T^\a=d\vt^\a+\G_\b{}^\a\wedge\vt^\b\,.\label{torscomp3} \ee The three
irreducible pieces of the torsion under $SO(1,3)$-decomposition are
displayed for later use in Tab.\ref{tab:irred}.

\begin{table}[h]
\tcaption{Irreducible decomposition of the torsion 
$T^\a={}^{(1)}T^\a+{}^{(2)}T^\a+{}^{(3)}T^\a$ under the Lorentz group 
$SO(1,3)$}
\bigskip
\label{tab:irred}
\begin{center}
\small
\begin{tabular}{||c|c|c|c||}\hline\hline
&&&\\
{} &{explicit expression} &{number of components} &{name}\\
&&&\\
\hline
&&&\\
{${}^{(2)}T^\a$} &{${1\over 3}\vt^\a\wedge(e_\b\rfloor T^\b)$} &{4} &{TRATOR}\\
&&&\\
${}^{(3)}T^\a$& $-{1\over 3}{}^*(\vt^\a\wedge{}^*(T^\b\wedge\vt_\b))$
&$4$&AXITOR\\
&&&\\
${}^{(1)}T^\a$&$T^\a-{}^{(2)}T^\a-{}^{(3)}T^\a$&16&TENTOR\\
&&&\\
\hline\hline
\end{tabular} 
\end{center}
\end{table}
\bigskip

\par
{}From the translational field strength $T^\alpha$, see (\ref{softcomm}) and 
(\ref{commstructure}), we come back to the translational gauge
potential $A^\a$.  According to the general gauge scheme, it will
couple to the energy-momentum current according to \be \TT_\a={{\d
    L}\over{\d A^\a}}\,. \label{ta} \ee The left hand side of
(\ref{tj}) suggests that the gauge process will replace the holonomic
coframe $dx^i$ by the potentials $A^\a$. Do the $A^\a$ have something
to do with an anholonomic coframe $\vt^\a$?  A consistent relation
between $T^\a, \vt^\a$, and $A^\a$ is established if we assume that
\be \vt^\a=\d^\a_i dx^i+A^\a\,.
\label{coframes} \ee Then we obtain \be dA^\a\,=\,d\vt^\a \,{\buildrel
  {*}\over{=}}\,T^\a \,.\label{fieldstrengths} \ee From
(\ref{fieldstrengths}) we get the `Bianchi' identity $d\,T^\a\stareq
0$. Comparison to the general Bianchi identity
$DT^\a=R_\b{}^\a\wedge\vt^\b$ indicates that the structures found so
far are part of a more general framework. This is, indeed, the case.
In order to derive GR from as little input as possible, we refrain at
this point from introducing additional fields, as for example an
independent linear connection $\G$. We collected the structures relevant
for the translational gauge scheme in Tab.\ref{tab:transl}.
 
\begin{table}[h]
  \tcaption{The relevant structures in a gauge approach of the four
    parameter translation group}
\label{tab:transl}
\bigskip\begin{center}
\small
\begin{tabular}{||c|c|c||}\hline\hline
{} & {}& {number of components}\\
\hline
&&\\
 conserved momentum current & $\TT_\a$ & $4\times 4$ \\
&&\\
 translational gauge potential & $\vt^\a\simeq A^\a$ (coframe)&$4\times 4$\\
&&\\
 translational field strength & $T^\a$ (torsion)& $4\times 6$ \\
&&\\
 1st Bianchi identity $(\Gamma\stackrel{*}{=} 0)$ & $d\,T^\a\;
{\buildrel {*}\over{=}} \;0 $& $4\times 4$\\
&&\\
\hline\hline
\end{tabular} 
\end{center}
\end{table}

\section{Theory of the translation gauge: From Einsteinian 
teleparallelism to GR}

As we argued in the last section, GR should be derivable from gauge
ideas in a fairly straightforward manner. The key to arrive at GR is
to start with the conserved momentum current and to gauge the
translation group that is connected with it.

\subsection{Translation gauge potential}

We commence with the very basics: Consider again a field theory in
Minkowski spacetime (pseudo-Cartesian coordinates) defined by the
Lagrangian $L_{\rm mat}=$ $L_{\rm mat}(\Psi, d\Psi)$. An explicit
dependence of $L$ on the coordinates $x^i$ is already forbidden by
rigid translational invariance, which we started from. The coordinates
enter more implicitly: The field $\Psi$ has to be expressed in terms
of differential forms in order to build the Lagrangian 4-form $L$ as
the appropriate integrand of the action. Therefore one needs the
differentials $dx^i$ as natural (or holonomic) basis for the physical
field.  They are invariant under rigid but not under soft
translations: \be \d x^i=\ve^i(x)\quad\lr\quad\d dx^i=d\ve^i\,.  \ee
Referring to the equivalence between active and passive
transformations, which is valid at this stage (see Sec.4.2), we view
in this approach the translations as passive transformations.  The
differentials $dx^i$ are no longer sufficient to build up an invariant
Lagrangian. According to the gauge principle, we have to introduce a
gauge potential $A^\a$ with transformation behavior $\d A^\a=-\d^\a_i
d\ve^i$ such that  \be \vt^\a:=\d^\a_i {\hat D}x^i :=\d^\a_i
dx^i+A^\a \ee transforms like \be \d \vt^\a =0\, .  \ee The
anholonomic one-form basis $\vt^\a$ serves as an appropriate form
basis for Lagrangians $L_{\rm mat}$ since it automatically
incorporates soft translational invariance. One usually refers to
$\vt^\a$, instead of $A^\a$, as the translational potential. Normally,
a distinction between $\vt^\a$ and $A^\a$ is not necessary, since
their field components differ just by a constant (1 or 0).

\subsection{Lagrangian}

The corresponding field strength $T^\a\stareq d\vt^\a$ can be used to
construct a kinematic supplementary term for $\vt^\a$ to the
Lagrangian.  The double role of $\vt^\a$ as both, a dynamical gauge
potential and an orthonormal frame (defining a new metric via
$g=o_{\a\b}\,\vt^\a\otimes\vt^\b$), explains the transition from
Minkowski space to a dynamical spacetime, which is due to
translational invariance. For the kinematic term we make the quadratic
ansatz $V=d\vt^\a\wedge H_\a$. What would be a good choice for $H_\a$?
Eyeing at Yang-Mills theory, we are tempted to put $H_\a={1\over
  2\ell^2}\,{}^* d\vt_\a$, with $\ell$ = Planck length. But we would
like to end up with a softly Lorentz invariant theory. The Lagrangian
$V={1\over 2\ell^2}\,{}^*d\vt^\a\wedge d\vt_\a$ is rigidly but not softly
Lorentz invariant, though. Note that this postulate of soft Lorentz
symmetry is not equivalent to a gauging of the Lorentz group! We won't
cure the lack of Lorentz invariance by the introduction of some new
gauge field $\G$ for the Lorentz field, but will use it just as a
criterion of choosing a good Lagrangian.

The most general term V quadratic in $d\vt^\a$ is obtained by choosing
$H_\a$ as \be H_\a={1\over 2\ell^2}\,{}^*\left(a_1\,{}^{(1)}d\vt_\a+
  a_2\,{}^{(2)}d\vt_\a+a_3\,{}^{(3)}d\vt_\a\right) \,, \label{Hdef}
\ee see also Mielke\cite{egg2,egg1}. The pieces ${}^{(I)}d\vt^\a$
correspond to the irreducible pieces ${}^{(I)}T^\a$ of the torsion,
compare Table \ref{tab:irred}: \bea {}^{(2)}d\vt^\a &:=& {1\over
  3}\vt^\a\wedge(e_\b\rfloor d\vt^\b)\,,\nonumber\\ {}^{(3)}d\vt^\a
&:=& -{1\over 3}{}^*\{\vt^\a\wedge {}^*(d\vt^\b\wedge\vt_\b)\}
\,,\nonumber \\ {}^{(1)}d\vt^\a &:=& d\vt^\a -
{}^{(2)}d\vt^\a-{}^{(3)}d\vt^\a\,.  \eea The postulate of soft
Lorentz invariance leads to a solution for the constant and real
parameters $a_I$ in the following way:

Infinitesimal
Lorentz rotations are expressed by $\d\vt^\a=\ve^\a{}_\b\,\vt^\b$ where
$\ve_{\a\b}=-\ve_{\b\a}$ are the antisymmetric Lorentz group
parameters.  It is easy to check that the gauge Lagrangian
$V=d\vt^\a\wedge H_\a$, with $H_\a$ given by (\ref{Hdef}), is
invariant under {\it rigid} Lorentz rotations, $\d V=0$.  The general
expression for $\d V$ reads \be \d V=\left({{\6 V}\over{\6\vt^\a}}-d{{\6
    V}\over{\6 d\vt^\a}}\right)\wedge \d\vt^\a +d\left({{\6 V}\over{\6
    d\vt^\a}}\wedge\d\vt^\a\right)\,.
\label{Vvar}
\ee Hence we have $\d V=0$ for rigid Lorentz rotations. However, for
{\it soft} Lorentz rotations with spacetime-dependent group
parameters $\ve_{\a\b}=\ve_{\a\b}(x)$, we get from (\ref{Vvar}) the
offending term \be \d_{\rm (soft)}V=d\ve^\a{}_\b\wedge{{\6 V}\over{\6
    d\vt^\a}}\wedge\vt^\b\,.  \ee In order to preserve Lorentz
invariance, this term has to be canceled, modulo an exact form. Using
the Leibniz rule, we obtain \be d\ve^\a{}_\b\wedge{{\6 V}\over{\6
    d\vt^\a}}\wedge\vt^\b= \ve^\a{}_\b\, d\left({{\6
      V}\over{\6\vt^\a}}\wedge\vt^\b\right) -d\left(\ve^\a{}_\b\, {{\6
      V}\over{\6\vt^\a}}\wedge\vt^\b\right)\,.  \ee The second term on
the r.h.s.\ is already exact. From the first term we get the condition
\be {{\6 V}\over{\6 d\vt}}{}_{[\a}\wedge\vt_{\b ]}={\rm exact}\; {\rm
  form} \ee for soft Lorentz invariance of $V$. We plug in the
explicit expression for $V$ and obtain, after some algebra, \bea {{\6
    V}\over{\6 d\vt}}{}{}_{[\a}\wedge\vt_{\b ]}&=&\left({1\over
    3}a_1-{1\over 3} a_3\right)d\h_{\a\b}-\left({2\over
    3}a_3+{1\over 3}a_1\right)d\vt_{[\a}\wedge\vt_{\b]}\nonumber \\ 
&+&\left({1\over 6}a_1+{1\over 6}a_2-{1\over
    3}a_3\right)\left(e_\g\rfloor d\vt^\g \right)\wedge \h_{\a\b}\,.
\eea The last two terms can be made vanishing by choosing \be
a_2=-2a_1\,,\qquad a_3=-{1\over 2}\,a_1\,. \ee Then we obtain \be {{\6
    V}\over{\6 d\vt}}{}_{[\a}\wedge\vt_{\b ]}={a_1\over
  2}\,d\h_{\a\b}\,.  \ee The constant $a_1$ can be absorbed by a
suitable choice of the coupling constant $\ell$ in $V$, see
(\ref{Hdef}). According to the usual conventions, we put $a_1=-1$,
i.e.\ $V$ is softly Lorentz invariant for the choice of parameters \be
a_1=-1\,,\qquad a_2=2\,,\qquad a_3={1\over 2}\,. \ee Hence \be
V_{||}={1\over 2\ell^2}\,d\vt^\a\wedge{}^*\left(-{}^{(1)}d\vt_\a+
  2\,{}^{(2)}d\vt_\a+{1\over 2}
  \,{}^{(3)}d\vt_\a\right)\,.\label{Vinv} \ee \bigskip

\begin{figure}
  \epsfbox[10 0 500 300]{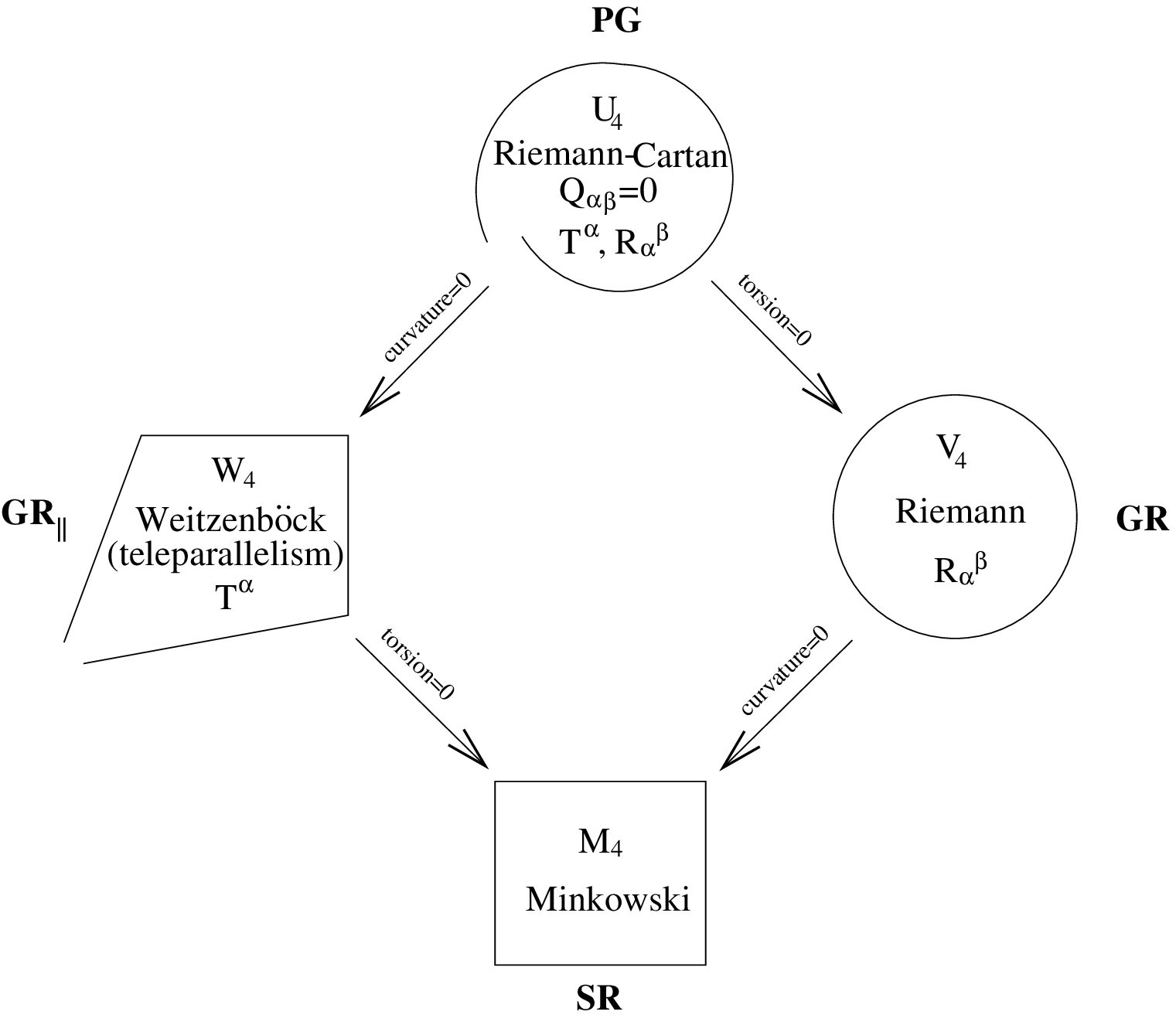}\bigskip \fcaption{A spacetime with a
    metric and a metric compatible connection (nonmetricity=0) is
    called a Riemann-Cartan space $U_4
$. It can either become a
    Weitzenb\"ock space $W_4$, if its curvature vanishes, or a Riemann
    space $V_4$, if the torsion happens to vanish. These different
    models of spacetime are the arenas for different gravitational
    theories.}
\label{fig:3}
\end{figure}

The total Lagrangian reads \be L_{\rm tot}=V_{||}+L_{\rm
  mat}(\Ps,d\Ps,\vt^\a)\,, \ee and the field equation ${{\d L_{\rm
      tot}}/{\d\vt^\a}}=0$ becomes \be
dH_\a-E_\a=\TT_\a\,,\label{fieldeq} \ee where, as before, $\TT_\a={{\d
    L_{\rm mat}}/{\d\vt^\a}}$ denotes the canonical energy-momentum
three-form and \bea E_\a&:=& (e_\a\rfloor d\vt^\b)\wedge H_\b-{1\over
  2}\,e_\a\rfloor(d\vt^\b\wedge H_\b)\nonumber \\&=& {1\over
  2}\,\left[(e_\a\rfloor d\vt^\b)\wedge H_\b-d\vt^\b\wedge(e_\a\rfloor
  H_\b)\right]\,,\eea the energy-momentum current of the gauge field.

\subsection{Transition to GR}

If the Lagrangian (\ref{Vinv}) is substituted into the field equation
(\ref{fieldeq}), then it can be seen that the antisymmetric piece of
the left hand side of (\ref{fieldeq}) vanishes, \be
\vartheta_{[\b}\wedge d H_{\a]}-\vartheta_{[\b}\wedge E_{\a]}=0
\,.\label{antisym}\ee Therefore the right hand side has to be
symmetric, too, i.e.\ only {\it scalar} matter fields or {\it gauge}
fields, such as the electromagnetic field, are allowed as material
sources, whereas matter carrying spin cannot be consistently coupled
in such a framework.
The existing nontrivial torsion, expressed by $d\vt^\a$, 
describes the nontrivial Riemannian 
geometry of spacetime. This is because we have tied the Christoffel
connection $\rG_{\a\b}$, which is determined by the metric, to the contortion 
tensor $K_{\a\b}$ by means of
the teleparallel condition $\Gamma_{\a\b}\stareq 0$:
\be
\rG_{\a\b}-K_{\a\b}=\G_{\a\b}\stareq 0\qquad\Longrightarrow\qquad
\rG_{\a\b}\stareq K_{\a\b}\,.   \label{tied}  
\ee
In other words: Meaningful teleparallel theories 
do {\it not} presuppose spinning matter as a source for nontrivial torsion, in 
contrast to what is sometimes stated in literature\cite{Kopz}.
\par
Doesn't all this look like general
relativity? The Levi-Civita (or Christoffel) connection, corresponding
to the metric $g=o_{\a\b}\,\vt^\a\otimes\vt^\b$, is given by \be
\rG{}_{\a\b}={1\over 2}\,\left(e_\a\rfloor d\vt_\b -e_\b\rfloor
  d\vt_\a- (e_\a\rfloor e_\b\rfloor d\vt_\g)\wedge\vt^\g\right)\,.
\label{rG} \ee The corresponding Riemannian curvature reads \be \rR
{}_{\a\b}=d\rG {}_{\a\b}-\rG {}_{\a\g}\wedge\rG {}^\g {}_\b\,.
\label{rR} \ee We use the last two equations to replace on the
Lagrangian level the variable $d\vt^\a$ by $\rG_{\a\b}$. Using
(\ref{rG}) and (\ref{rR}), one can prove the remarkable identity \be
{1\over 2}\, \rR {}^{\a\b}\wedge\h_{\a\b}-{\ell^2}\,V
=d(\vt^\a\wedge {}^* d\vt_\a)\,, \label{identity} \ee with $V$ given
by (\ref{Vinv}).  Therefore one finds that the kinematic term $V$,
with the above choice of parameters $a_I$, is equivalent to the
Hilbert-Einstein action modulo an exact term.  Replacing $V$ in the
action $S$ by means of (\ref{identity}) leads, via ${{\d L_{\rm
      tot}}/{\d \vt^\a}}=0$, to Einstein's equation \be \rE
{}_\a:={1\over 2}\,\eta_{\a\b\g}\wedge\rR {}^{\b\g} ={\ell^2}\,
\TT_\a\,.  \ee In such a way, we arrive at GR in its original
form. Shifting back and forth from the variable pair $(\vt^\a,
\rG_\a{}^\b)$ to $(\vt^\a,d\vt^\a)$ means shifting back and forth from
original GR to its teleparallel equivalent GR$_{||}$. This is
displayed in Fig.\ref{fig:3}: A general PG induces a Riemann-Cartan
space $U_4$ with nonvanishing torsion and curvature.  Such a $U_4$ can
be reduced to either a Weitzenb{\"o}ck space $W_4$ (curvature $=0$) or
a Riemann space $V_4$ (torsion $=0$), i.e.\ to the geometries induced
by GR$_{||}$ or GR, respectively.
\par
We wish to point out that both theories can be obtained as special
cases within the framework of the Poincar\'e gauge theory PG, see our
Physics Reports\cite{PRs}, the literature given there, and the work of 
Pascual-S\'anchez\cite{PS2,PS3}.

\section{Gauging of the affine group $R^{4}\semidirect GL(4,R)$}

We recognized that the gauging of the translations yields a theory
which, for spinless matter and for electromagnetism, that is, for
symmetric energy-momentum currents, is equivalent to GR. Thus we have
a new understanding of Einstein's theory. Why should we generalize
GR$_{||}$ if it is consistent with experiment? Three somewhat
interrelated arguments come to mind:
\begin{itemize}
\item The translations represent a subgroup of the Poincar\'e group.
  Only to gauge the translations and to leave the Lorentz subgroup of
  the Poincar\'e group untouched, would seem unnatural. This argument
  is all the more convincing, since the {\it semi-}direct product
  structure of the Poincar\'e group interrelates its two mentioned
  subgroups stronger than it were the case for a direct product.
\item A Weitzenb\"ock spacetime is a degenerate Riemann-Cartan space.
  The gauge arguments against the rigidity of a teleparallelism were
  already advanced by Weyl in the twenties against Einstein's
  corresponding theory. 
\item The translational gauge procedure, as it is obvious from the
  field equation (\ref{fieldeq}) with (\ref{Vinv}), works only for
  spinless matter and for electromagnetism, since the field equation
  is symmetric, see (\ref{antisym}), and supplies only 10 independent
  components.
\end{itemize}
We conclude that a gauging of the whole $4+6$ parameter Poincar\'e group
is mandatory.  The theory which emerges, the Poincar\'e gauge theory (PG),
is formulated in a Riemann-Cartan spacetime, the Einstein-Cartan
theory (EC) being a degenerate subcase of it. And the EC is a viable
gravitational theory!
\par
In order to see the built-up of the different structures of spacetime
more clearly, we prefer to gauge\cite{egg0} immediately the $4+16$
parameter {\it affine} group\cite{egg0} $A(4,R)= R^4\semidirect
GL(4,R)$ --- which lacks a metric structure altogether --- and to
introduce the metric {\it subsequently}. Thus we arrive at the
metric-affine gauge theory (MAG) which encompasses the PG as a
subcase. Symbolically, we may write \be {\rm MAG}/_{\rm
  nonmetricity\,=\,0}\;=\;{\rm PG}\,.\ee
\par

The actual gauging of the affine group was done in our Physics
Reports\cite{PRs}, and we follow the presentation given there. A short
outline of these results, see Secs.\ 3.1, 3.2, 3.3, loc.\ cit., will
be given here. We start then in the flat $n$--dimensional affine space
$R^n$. The rigid affine group $A(n,R):= R^{n}\semidirect$ $GL(n,R)$ is
the semidirect product of the group of $n$-dimensional {\it
  translations} and $n$-dimensional {\it general linear
  transformations}. This transformation group, cf.\ ref.(\cite{Ko13})
p.\ 27, acts on an affine $n$--vector $x=\{x^{\alpha}\}$ according to
\begin{equation} 
x\qquad \rightarrow \qquad x'=\Lambda\, x +
\tau \,,
\end{equation} 
where $\Lambda=\{\Lambda^{\alpha}{}_{\beta}\}\in GL(n,R)$ and 
$\tau=\{\tau^{\alpha}\} \in R^n$.
Thus it is a generalization of the Poincar\'e group 
$P:= R^{4}\semidirect$ $SO(1,3)$,
with the pseudo--orthogonal group $SO(1, n-1)$ being replaced by the
general linear group $GL(n, R)$. It is convenient to work 
with a
M\"obius type representation for which we take the same
symbol $A(n,R)$: It is that subgroup of $GL(n+1,R)$ which
leaves the $n$-dimensional hyperplane ${\buildrel =\over
R}{}^{n}:=\left\{{\buildrel =\over x} =
\pmatrix{x\cr 1\cr}\in R^{n+1}\right\}$ 
invariant:
\begin{equation}A(n,R) = \left\{ \pmatrix{\Lambda& \tau\cr 0& 1\cr}\in
GL(n+1,R)\left\vert \; \Lambda\in GL(n,R), \; \tau\in
R^n\right.\right\}.
\end{equation}
Thus, by an affine transformation, we obtain 
\be
{\buildrel =\over x}'
=\pmatrix{\Lambda& \tau\cr 0& 1\cr}\pmatrix{x\cr
1\cr}=\pmatrix{\Lambda x+\tau\cr 1\cr}\,,
\ee 
as is required for the action of the affine group on the flat affine space.
The Lie algebra $a(n,R)$ consists of the generators 
$P_{\gamma}\,$, representing $n$-dimensional translations, and the 
$L^{\alpha}{}_{\beta}\,$, 
which span the 
Lie algebra $gl(n,R)$ of $n$-dimensional linear transformations. 

In a matrix representation we can write the
affine {\it gauge} group as
\begin{equation}{\cal A}(n,R) = \left\{ \pmatrix{\Lambda(x)& \tau (x)\cr
0& 1\cr}\left\vert \; \Lambda(x)\in {\cal G}{\cal L}(n,R), \quad 
\tau (x)\in {\cal T}(n,R)\right.\right\}\,.\label{3.2.1}
\end{equation}
Having already had some experience with the Yang-Mills type gauge approach, 
we are aware of the need of introducing a gauge potential in order to
step from rigid to soft group invariance. Accordingly, by gauging the 
affine group, the softening of the affine group transformations is
accompanied by the introduction of the 
{\it generalized affine} connection ($\rightarrow$ potential)
\begin{equation}{\buildrel\approx\over\Gamma} =\pmatrix{\Gamma^{(L)}& 
\Gamma^{(T)}\cr
                           0& 0\cr}
=\pmatrix{\Gamma^{(L)}_{\alpha}{}^{\beta}\,L^{\alpha}{}_{\beta}&
\Gamma^{(T)\alpha}\,P_{\alpha}\cr
                           0& 0\cr}\,.\label{3.2.2}\end{equation}
It is a one--form ${\buildrel\approx\over\Gamma}=
{\buildrel\approx\over\Gamma}_{i}\, dx^{i}$ and 
transforms inhomogeneously under
an affine gauge transformation,
\begin{equation}{\buildrel\approx\over\Gamma}\quad
{\buildrel {A^{-1}(x)}\over\longrightarrow}\quad
{\buildrel\approx\over\Gamma}{}'=
A^{-1}(x)\,{\buildrel\approx\over\Gamma}\, A(x) + A^{-1}(x)dA(x)\;
,\qquad A(x)\in {\cal A}(n,\, R),\label{3.2.3}
\end{equation} 
where the transformation is formed with respect to the group element
\begin{equation}A^{-1}(x)=\pmatrix{\Lambda^{-1}(x)& -\Lambda^{-1}(x)\tau (x)\cr
                        0& 1\cr}\; . \label{3.2.4} \end{equation}
The corresponding affine curvature is given by 
\begin{equation}{\buildrel\approx\over R}:= d{\buildrel\approx\over\Gamma} +
{\buildrel\approx\over\Gamma}\wedge {\buildrel\approx\over\Gamma} =
\pmatrix{R^{(L)}&R^{(T)}\cr 0& 0\cr}=
\pmatrix{d\Gamma^{(L)} +\Gamma^{(L)}\wedge\Gamma^{(L)} & 
d\Gamma^{(T)}+ \Gamma^{(L)}\wedge \Gamma^{(T)}\cr 0& 0\cr}\,. \label{3.2.5}
\end{equation} 
It transforms covariantly under the affine gauge group:
\begin{equation}{\buildrel\approx\over R}\quad{\buildrel 
{A^{-1}(x)}\over\longrightarrow}\quad {\buildrel\approx\over R}'= A^{-1}(x)
\,{\buildrel\approx\over R}\, A(x)\,.\label{3.2.6}\end{equation} 
The exterior covariant derivative ${\buildrel\approx\over D}:= d +
{\buildrel\approx\over \Gamma}\wedge$ acts on an affine $p$--form
${\buildrel\approx\over\Psi} = \pmatrix{\Psi\cr 1\cr}$ as follows: 
\begin{equation}{\buildrel\approx\over D}\,{\buildrel\approx\over\Psi}=
\pmatrix{d\Psi+\Gamma^{(L)}\wedge\Psi 
+\Gamma^{(T)}\cr
0\cr}=\pmatrix{D\Psi+\Gamma^{(T)}\cr 0\cr}\,
.\label{3.2.7}\end{equation} 

After substitution of (\ref{3.2.2}) and (\ref{3.2.4}), 
the inhomogeneous transformation 
law (\ref{3.2.3}) splits into
\begin{equation}\Gamma^{(L)}\quad{\buildrel {A^{-1}(x)}\over
\longrightarrow}\quad\Gamma^{(L)}{}'= \Lambda^{-1}(x)\,\Gamma^{(L)}\, 
\Lambda(x) + \Lambda^{-1}(x)d\Lambda(x)\;, \label{3.2.8}
\end{equation}
and 
\begin{equation}\Gamma^{(T)}\quad{\buildrel {A^{-1}(x)}\over
\longrightarrow}\quad\Gamma^{(T)}{}'= \Lambda^{-1}(x)\, \Gamma^{(T)}+
 \Lambda^{-1}(x)D\tau (x)\; .\label{3.2.9}
\end{equation} 
The soft translations
$\tau (x)$ automatically drop out in (\ref{3.2.8}) due to the one--form
structure of $\Gamma^{(T)}$. Thereby (\ref{3.2.8}) acquires the conventional
transformation rule (with the exterior derivative $d$) of a
Yang--Mills--type connection for ${\cal G}{\cal L}(n,R)$. Thus we can
identify $\Gamma^{(L)}=\Gamma= \Gamma_{\alpha}{}^{\beta}\,
L^{\alpha}{}_{\beta}$ with the linear connection and are basically
done with the gauging of the linear part of the affine group. 
 
\begin{table}[h]
  \tcaption{Gauge fields in metric-affine gauge theory MAG}
\label{arena}
\bigskip
\begin{center}
\small
\begin{tabular}{||c|c|c||}\hline\hline
{Potential} & {Field strength}& {Bianchi identity}\\
\hline
&&\\
 metric $g_{\a\b}$ & $Q_{\a\b}=-Dg_{\a\b}$ & $DQ_{\a\b}=2R_{(\a}{}^\mu
\,g_{\b)\mu}$ \\
&&\\
 coframe $\vt^\a$ & $T^\a=D\vt^\a$ & $DT^\a=R_\mu{}^\a\wedge\vt^\mu$ \\
&&\\
 connection $\G_\a{}^\b$ & $R_\a{}^\b=d\G_\a{}^\b-\G_\a{}^\mu\wedge\G_\mu{}^\b$
& $DR_\a{}^\b=0$ \\
&&\\
\hline\hline
\end{tabular} 
\end{center}
\end{table}

Things are more involved for the translational part, though. From what
we have learned by gauging the translation group we expect
$\Gamma^{(T)}$ to be related to the coframe $\vartheta
:=\vartheta^{\alpha}\, P_{\alpha}$, i.e.\ to a one--form with values
in the Lie algebra of $R^n$.  But due to the covariant exterior
derivative term $D\tau (x):= d\tau (x)+ \Gamma^{(L)}\, \tau (x)$ in
(\ref{3.2.9}), the translational part $\Gamma^{(T)}$ does not
transform as a covector, as is required for the coframe $\vartheta$.
To get a correspondence between $\Gamma^{(T)}$ and $\vartheta$, we
introduce a vector (vector--valued zero--form)
${\buildrel\approx\over\xi} = \pmatrix{\xi\cr 1\cr}=
\pmatrix{\xi^{\alpha}\, P_{\alpha}\cr 1\cr}$ which transforms as
${\buildrel\approx\over\xi}{}'=A^{-1}(x)\,
{\buildrel\approx\over\xi}$, i.e.\ as
\begin{equation}\xi\quad {\buildrel {A^{-1}(x)}\over\longrightarrow}\quad  
\xi'=\Lambda^{-1}(x)\,\left(\xi- \tau (x)\right)\label{3.2.10}\end{equation}
under an active affine gauge transformation.  Then we define
\begin{equation}\vartheta :=\Gamma^{(T)} + D\xi \label{3.2.11}\end{equation}
which transforms as a vector--valued
one--form under the ${\cal A}(n,R)$, as required:
\begin{equation}\vartheta\quad {\buildrel {A^{-1}(x)}\over
\longrightarrow}\quad \vartheta'=\Lambda^{-1}(x)\, \vartheta\,.\label{3.2.12}
\end{equation}
The geometric and physical meaning of the relation (\ref{3.2.11}),
especially the role of the field $\xi$, is perhaps not completely
satisfactorily clarified, yet. For an expanded discussion of this
point we refer the reader again to ref.(\cite{PRs}), see also
Mistura\cite{Mistura}. The translational piece of the affine
connection, namely $\Gamma^{(T)}$, could play a decisive role in the
gravitationally induced phase factor of a matter wave, as was pointed
out by Morales-T\'ecotl et al.\cite{Alfredo2'}. Here, we confine ourselves to
require the condition
\begin{equation}
D\xi=0\,
\end{equation}
to hold. 
Then the generalized affine connection
${\buildrel\approx\over\Gamma}$ on the affine bundle $A(M)$ reduces to
the {\it Cartan connection}
\begin{equation}
{\buildrel =\over\Gamma} =\pmatrix{\Gamma^{(L)}&  \vartheta\cr
                           0& 0\cr} 
\end{equation}
on the bundle $L(M)$ of linear frames. Due to (\ref{3.2.12}), this is not
anymore a connection in the usual sense. We are thus left with the 
potentials $(\vt^\a, \G_\a{}^\b)$ and their corresponding field strengths
$(T^\a, R_\a{}^\b)$. 

For this gauging of the affine group no metric was necessary. If
additionally a {\it metric} is given, we recover the metric--affine
geometrical arena of Sec.\ 2.4, as is summarized in Table \ref{arena}.
 
We close this geometric section by introducing a further geometric
ingredient, the ``$\eta$-bases'', which span the graded algebra of
dual exterior forms on each cotangent space and which turn out to be
quite useful in practical calculations.  First, owing to the existence
of a metric, we can define the scalar density $\sqrt{|\det
  g_{\mu\nu}|}$ and the familiar Hodge star operator $\hodge\,$. The
Hodge star operator maps a $p$--form $\Psi$ into an $(n-p)$--form
$^*\Psi$ by means of the explicit formula \be ^*\Psi:=
{1\over{(n-p)!\, p!}}\, \sqrt{|\det g_{\mu\nu}|}\,
g^{\alpha_1\gamma_1}\cdots g^{\alpha_p\gamma_p}\,
\epsilon_{\alpha_1\cdots\alpha_p\beta_1\cdots\beta_{n-p}}\,
\Psi_{\gamma_1\cdots\gamma_p}\,
\vartheta^{\beta_1}\!\wedge\cdots\wedge\vartheta^{\beta_{n-p}}\,.\ee

Now we define the $g$-volume
element $n$--form
\be \eta:=\sqrt{|\det g_{\mu\nu}|}\,\vta^{\hat{1}}\wedge\cdots\wedge
\vta^{\hat{n}}={1\over n!}\,\sqrt{|\det g_{\mu\nu}|}\,\epsilon_{
\a_1\cdots\a_n}\vta^{\a_1}\wedge\cdots\wedge\vta^{\a_n}={}^*1\,,\ee
dual to the unit zero--form. Picking a (pseudo-)orthonormal 
positively oriented coframe
$\stackrel{\circ}{\vta}{}^\a$, the $g$-volume element simplifies to
\begin{equation} \eta  = \,\stackrel{\circ}{\vartheta}{} 
^{\hat{1}}\wedge \cdots \wedge\stackrel{\circ}{\vartheta}{}^{\hat{n}} 
\,.\end{equation} 

Having this $n$--form at our disposal, we can successively 
contract it by means of the frame $e_\a$, thereby arriving at
an $(n-1)$--form, an $(n-2)$--form, etc., until we terminate the
series with a zero--form:
\begin{eqnarray}\label{etabasis}
\eta_{\alpha_{1}} & :=& e_{\alpha_{1}} \rfloor \eta =
{1\over{(n-1)!}}\, \eta_{\alpha_{1} \alpha_{2} \cdots \alpha_{n}}\,
\vartheta^{\alpha_{2}} \wedge \cdots \wedge \vartheta^{\alpha_{n}}=
\,^{*}\!\vartheta_{\alpha_{1}}\,, \nonumber \\
 \eta _{\alpha_{1} \alpha_{2}}& :=&
e_{\alpha_{2}} \rfloor \eta _{\alpha_{1}}
= {1\over{(n-2)!}}\, \eta _{\alpha_{1} \alpha_{2} \alpha_{3}\cdots 
\alpha_{n}}\,
\vartheta ^{\alpha_{3}} \wedge \cdots\wedge
\vartheta ^{\alpha_{n}}=\,^{*}\!(\vartheta_{\alpha_{1}}\wedge
\vartheta_{\alpha_{2}}) \,, \nonumber \\
 \quad &\vdots & \\ \eta _{\alpha_{1}\cdots\alpha_{n}}\, &:=&
e_{\alpha_{n}} \rfloor \cdots\rfloor e_{\alpha_{1}}\rfloor\eta =\,
^{*}\!(\vartheta_{\alpha_{1}}\wedge\cdots\wedge
\vartheta_{\alpha_{n}}) \,.\nonumber
\end{eqnarray}
Thus, in four dimensions, which we are concentrating on, we have an
$\eta$-basis at our disposal
$\left(\eta,\,\eta_\a,\,\eta_{\a\b},\,\eta_{\a\b\gamma},
  \,\eta_{\a\b\gamma\delta}\right)$ that we will meet in numerous
applications.

\section{Field equations of metric--affine gauge theory (MAG)}

The gauge procedure of the affine group led to the identification of
the gauge potentials of the MAG, see Table 5. The material currents
are then expected to couple to these potentials in the Yang-Mills type
fashion, as is indicated in Fig.\ 1, and as we know already from the
material energy-momentum current and the coframe according to the line
after (\ref{fieldeq}). If $\Psi$, as a $p$-form, represents a matter
field (fundamentally a representation of the $SL(4,R)$ or of some of
its subgroups\cite{Ne2,Ne3}), its first order Lagrangian $L$ will be
embedded in metric-affine spacetime by the minimal coupling procedure,
that is, exterior covariant derivatives feature in the kinetic terms
of the Lagrangian instead of only exterior ones. Then the material
currents are defined as follows: \be \sigma^{\alpha\beta} :=
2\,{{\delta L}\over{\delta g_{\alpha\beta}}}\,, \qquad
\Sigma_{\alpha}:= {{\delta L}\over{\delta\vartheta^{\alpha}}}\,,
\qquad \Delta^{\alpha}{}_{\beta} := {{\delta
    L}\over{\delta\Gamma_{\alpha}{}^{\beta}}} \,.\label{mcurrents} \ee

{}From GR we expect that $\sigma^{\a\b}$ is the metric (and symmetric)
energy-momentum current of matter (`Hilbert current'), whereas Secs.\ 
4 and 5 lead us to the believe that $\Sigma_\a$ should be the
corresponding canonical energy-momentum current (`Noether current').
The canonical hypermomentum current $\Delta^\a{}_\b$, which couples to
the linear connection, can be decomposed according to
\ba\Delta_{\alpha\beta}\qquad &=&\qquad
{\tau_{\alpha\beta}}\qquad+\qquad {1\over
  4}\,g_{\alpha\beta}\;\Delta\qquad+\qquad
\,{\buildrel\frown\over{\Delta\quer}}_{\alpha\beta}\\ &\sim&\>{\rm
  spin}\> {\rm current}\,\oplus\,{\rm dilation}\> {\rm
  current}\,\oplus\, {\rm shear} \> {\rm current}\,, \nonumber\ea with
$\tau_{\alpha\beta}:= \Delta_{[\alpha\beta]}$ as (dynamical) spin
current, $\Delta:= \Delta^\gamma{}_\gamma\,,\label{dilation}$ as
dilation current, and
${\buildrel\frown\over{\Delta\quer}}_{\alpha\beta}$ as symmetric and
tracefree shear current.

If one applies the Noether theorem to the material Lagrangian, then,
$\Sigma_\a$, $\tau_{\a\b}$, and $\Delta$ can be identified with the
special-relativistic canonical Noether currents of energy-momentum,
spin, and dilation, respectively. This leaves only the shear current
${\buildrel\frown\over{\Delta\quer}}_{\alpha\beta}$, see our earlier
work\cite{Gr8} for a detailed discussion, as a concept that is a bit
more remote from direct observation than the other currents.  In any
case, the definitions (\ref{mcurrents}) and their physical
interpretations are completely justified by these considerations.
\par
Adding to the matter Lagrangian a metric-affine first order gauge
Lagrangian $V$, \be W = \int\left[V(g_{\alpha\beta},
  \vartheta^{\alpha}, Q_{\alpha\beta}, T^{\alpha},
  R_{\alpha}{}^{\beta}) + L(g_{\alpha\beta}, \vartheta^{\alpha},\Psi ,
  D\Psi)\right]\,,\label{action}\ee and applying the action principle,
we find\cite{PRs} the matter and the gauge field equations (of
Yang-Mills type): \ba D\left({\partial L\over\partial (D\Psi)
    }\right)-(-1)^p\, {{\partial L}\over{\partial\Psi}} &=&\quad\,
0\,,\quad\qquad\qquad {\rm (MATTER)}\\ & &\nonumber\\ 
D\left(2\,{\partial V \over\partial Q_{\alpha\beta}
    }\right)+2\,{\partial V \over\partial g_{\alpha\beta} }
&=&-\,\sigma^{\alpha\beta}\,, \qquad\,\qquad {\rm (ZEROTH)}
\label{zeroth}\\
& &\nonumber\\
D\left({\partial V\over\partial T^\alpha }\right)
+{\partial V\over\partial\vartheta^\alpha}& =&  
-\,\Sigma_{\alpha}\,, \qquad\>\;\qquad {\rm (FIRST)}\label{first}\\
& &\nonumber\\
D\left({\partial V\over\partial R_\a{}^\b}\right)+\vartheta^\a\wedge
{\partial V\over\partial T^\b}+2\,g_{\b\gamma}\,  {\partial V\over\partial 
Q_{\a\gamma}}
 &=&  -\,\Delta^{\alpha}{}_{\beta}\,. \qquad\qquad\, {\rm (SECOND)}
\label{second}\ea  
In SECOND a Noether identity for $V$ has already been employed.
Analogously, in FIRST, the canonical energy-momentum of the
translational gauge potential $\vartheta^\a$ can be expressed
explicitly as \be {{\6 V}\over{\6 \vt^\a}}\;=\;e_\a\rfloor V
-(e_\a\rfloor T^\b)\wedge {{\6 V}\over{\6 T^\b}} -(e_\a\rfloor
R_\b{}^\g)\wedge{{\6 V}\over{\6 R_\b{}^\g}}- (e_\a\rfloor
Q_{\b\g})\,{{\6 V}\over{\6 Q_{\b\g}}}\,.\label{gaugeem} \ee This
structure is known from Minkowski's energy-momentum tensor of the
Maxwellian field. It is interesting to note that, provided SECOND is
fulfilled, FIRST and ZEROTH are equivalent, i.e., one of them is
redundant. It is for this reason that we abstain, for $\partial
V/\partial g_{\a\b}$, to display a formula similar to
(\ref{gaugeem}).

The field equation of a Yang-Mills theory reads
$\stackrel{A}{D}(\partial V_{\rm YM}/\partial F)=-J$. The field
equations ZEROTH, FIRST, and SECOND are of this type. However, because
of the universality of `external' interactions (gravitation),
additional {\it tensor-}valued {\it gauge} currents ($\partial
V/\partial\vartheta^\a$ etc.) surface in the field equations. This is
the distinguishing feature of MAG as compared to gauge theories of
internal groups ($U(1),\,SU(2),\dots$). In Yang-Mills theory the
non-linearity of the gauge field is hidden in the non-tensorial pieces
of the gauge covariant exterior derivative $\stackrel{A}{D}$ occurring
on the left hand side $\stackrel{A}{D} (\partial V_{\rm YM}/\partial
F)$ of the Yang-Mills equation. In gravitational gauge theory, there
are {\it additional} non-linearities, besides those in
$\stackrel{\Gamma}{D}$, namely the ones represented by (\ref{gaugeem})
etc.. This result is non-trivial as a simple argument will show:
\par
Suppose we had a Hilbert-Einstein gauge Lagragian linear in the
curvature $R_\a{}^\b$. Then in FIRST the leading term in differential
order on its left hand side will vanish and we are left with
(\ref{gaugeem}). The surviving terms are \be {{\6 V_{\rm HE}}\over{\6
    \vt^\a}}\;\sim\;e_\a\rfloor V_{\rm HE} -(e_\a\rfloor
R_\b{}^\g)\wedge{{\6 V_{\rm HE}}\over{\6
    R_\b{}^\g}}\;\sim\;-\,\Sigma_\a\,.\label{gaugeHE} \ee Since ${\6
  V_{\rm HE}}/{\6 R_\b{}^\g}$ will be a constant, we recover the
Einstein three-form (corresponding to the Einstein tensor in Ricci
calculus) from this equation, giving substance to Schr\"odinger's
dictum\cite{Sc7} that the left hand side of Einstein's equation is, in
some sense, the gravitational energy-momentum tensor. Consequently,
Einstein's field equation of GR is encapsulated in $\6 V/\6
\vartheta^\a$ of FIRST and, as such, has a distinctive anti-Yang-Mills
flavor. In contrast, the Einsteinian teleparallelism GR$_{||}$, with
its torsion-square Lagrangian$V_{||}$, picks up an essential piece
from the proper Yang-Mills term of FIRST, \be D\frac{\6 V_{||}}{\6
  T^\a} +e_\a\rfloor V_{||}-(e_\a\rfloor T^\b)\wedge\frac{\6
  V_{||}}{\6 T^\b}\;\sim\;-\,\Sigma_\a\,,\label{tele}\ee compare
(\ref{fieldeq}).  It has much more of the Yang-Mills spirit than GR
has -- and this is the reason why GR$_{||}$ turned up when we gauged
the translations in Secs.\ 4 and 5.
\par
If, within our formalism, one desires to correctly derive the field
equation for GR, see (\ref{gaugeHE}), and for GR$_{||}$, see
(\ref{tele}), then one has to put on {\it Lagrange multipliers}. In
the case of (\ref{gaugeHE}) they have to kill nonmetricity and torsion
and for (\ref{tele}) to remove nonmetricity and curvature.  The
details have been worked out in our earlier review\cite{PRs}, see also
Kopczy{\'n}ski\cite{Ko7}. {\bf MAG} (metric-affine gauge theory, the
general framework), {\bf PG} (Poincar\'e gauge theory, vanishing
nonmetricity, hence in a Riemann-Cartan spacetime), {\bf EC}
(Einstein-Cartan theory, the PG with the curvature scalar as
gravitational Lagrangian), {\bf GR$_{||}$} (Einsteinian
teleparallelism, vanishing nonmetricity, vanishing curvature, hence in
a Weitzenb\"ock spacetime, specific torsion square Lagrangian), and
{\bf GR} (general relativity, vanishing nonmetricity, vanishing
torsion, hence in a Riemannian space, curvature scalar as Lagrangian)
are different (sub-)cases of this general scheme.

\par A further remark is in order: Gauging the affine group yields the
gauge potentials $(\vartheta^\alpha,\,\Gamma_\alpha{}^\beta)$, see
Table 5. If a metric exists on top of that linearly connected
manifold, then a further independent geometrical field variable is at
hand. Following Trautman\cite{Trautman}, we are taking
$(g_{\alpha\beta},\,\vartheta^\alpha,\, \Gamma_\alpha{}^\b)$ as
independent variables in the action (\ref{action}).  Because of the
redundancy of ZEROTH or FIRST, {\it provided} SECOND is fulfilled, one
could argue that one should drop, say, the coframe as independent
variable, as done by Tucker\footnote{We are grateful to Robin W.\ 
  Tucker (Lancaster) for an interesting discussion on this question.}
$\,$ and Wang\cite{Tucker1}, for example. In the earlier metric-affine
unified field theories \`a la Einstein\cite{meaning}(App.2) and
Schr\"odinger\cite{Sc7}, the coframe didn't even show up since they
worked in a holonomic formalism (unsuitable for representing
fermions). Because of various arguments, however, we feel more
comfortable with our procedure: (i) Both energy-momentum currents, the
Hilbertian $\sigma^{\alpha\beta}$ and the Noetherian $\Sigma_\a$, have
a good and direct physical interpretation in SR.  Dropping a gauge
variable means dropping one of these useful quantities as a
fundamental current. (ii) {\it A posteriori}, we do find the identity
causing the redundancy mentioned. This seems safer than to assume
something to that extend {\it a priori}. (iii) If we dropped the
metric as an independent variable, for example, then we would have to
take {\it orthonormal} frames as field variables, giving away a piece
of freedom which we had earlier in being able to work with whatever
frame we liked, be it orthonormal or oblique.
\par
Let us finally remind ourselves: As soon as the gauge Lagrangian $V$
is specified explicitly, we can find the field equations ZEROTH,
FIRST, and SECOND by sheer partial differentiation. Hence the using of
the general form of the field equations may save a lot of work.
\par

\section{Model building: Einstein-Cartan theory and beyond}

The missing piece within the framework that we finally established in
Sec.\ 7 is the gauge field Lagrangian $V(g_{\a\b},
\vartheta^\a,Q_{\a\b},T^\a,R_\a{}^\b)$. The hope is that the model,
with a suitably chosen $V$ -- perhaps combined with some symmetry
breaking mechanism which, for example, reduces the linear group
$GL(4,R)$ to the Lorentz group $SO(1,3)$ -- can be consistently
quantized. For 1+1 dimensional curvature square
models\cite{egg3,Yu1,Yu4} successful quantization methods are already
available, see Kloesch and Strobl\cite{Strobl1} and the literature
given there.
\par
In testing a new framework, one first wants, in some limit, to recover
old ground where one feels at home. In our case this is GR. There are at
least two ways of how to achieve this: One can choose the curvature
scalar \`a la Hilbert (in exterior calculus: the corresponding
four-form $R^{\a\b}\wedge\eta_{\a\b}$ \`a la Trautman\cite{Trautman})
and take Lagrange multipliers for extinguishing nonmetricity and
torsion: \be V_{\rm
  GR}=-\frac{1}{2\ell^2}\,R^{\a\b}\wedge\eta_{\a\b}+\frac{1}{2}
\,Q_{\a\b}\wedge{}^{(1)}\lambda^{\a\b}+T^\a\wedge{}^{(2)}\lambda_\a\,.
\label{VGR}\ee
Note that $R_\a{}^\b$ is the curvature tensor of the independent field
variable $\Gamma_\a{}^\b$, for the $\eta$'s see (\ref{etabasis}). 

\subsection{Einstein-Cartan theory EC}

A second
way is to start with the Einstein-Cartan Lagrangian (here with
cosmological constant) \be V_{\rm EC}=
-\frac{1}{2\ell^2}\,R^{\a\b}\wedge\eta_{\a\b}+
\frac{1}{2}\,\Lambda\,\eta+\frac{1}{2}
\,Q_{\a\b}\wedge\lambda^{\a\b}\,,\ee and to derive the
corresponding field equations: \bea {1\over 2}\,\eta_{\a\b\g}\wedge
R^{\b\g}+ \Lambda\,\eta_\a &=& \ell^2\, \Sigma_\a\,,\label{ECfirst} \\ 
{1\over 2}\,\eta_{\a\b\g}\wedge T^\g \!\,\;\quad &=& \ell^2\,\tau_{\a\b}
\,.\label{ECsecond}\eea In components in a holonomic frame they read:
\bea Ric_{\,ij}-{1\over 2}\,g_{ij}Ric_{\,k}{}^k +\Lambda\,
g_{ij}&=&\ell^2\, \Sigma_{ij}\,, \\ 
T_{ij}{}^k+2\,\d^k_{[i}\,T_{j]l}{}^l\;\quad &=& \ell^2\,\tau_{ij}{}^k\,.
\eea We recover GR for vanishing spin $\tau_{\a\b}=0$. In this context
only the vanishing of nonmetricity had to be assumed. The vanishing of
torsion, for spinless matter, was the result of the second field
equation (\ref{ECsecond}). We will see below, in Sec.\ 8.2, that also
the last Lagrange multiplier can be abandoned if one amends the
gravitational part of the Lagrangian with a piece quadratic in the
Weyl one-form. 
\par

Basically, the two EC field equations (\ref{ECfirst}) and
(\ref{ECsecond}) are {\it first order} partial differential equations
in $\Gamma_\a{}^\b$ and $\vartheta^\a$, with a spin fluid as source,
see ref.(\cite{Yu2}). As we explained already in the last Sec.\ 7, this
is the anti-Yang-Mills flavor of GR or EC caused by the absence of an
explicit torsion piece in the Lagrangian. A physical consequence is
that in EC we have the usual gravitational interaction of the
Newton-Einstein type plus a very weak (non-propagating) spin contact
interaction. Up to the fifties, {\it weak} interaction was also
thought to be of contact type (in fact, of a vector-axial vector
type). Later, following gauge ideas, the short-range intermediate
$W$-boson (and the $Z$) were postulated that made the weak interaction
propagate.

\subsection{Poincar\'e gauge theory PG, the quadratic version}

The simplest appearance of explicit torsion pieces is that in the
GR$_{||}$ Lagrangian\cite{Haynew,Erice} studied in Sec.\ 5: \be
V_{||}=-{1\over
  {2\ell^2}}T^\a\wedge{}^*\left(-{}^{(1)}T_\a+2{}^{(2)}T_\a +{1\over
    2}{}^{(3)}T_\a\right)+\frac{1}{2}
\,Q_{\a\b}\wedge{}^{(1)}\lambda^{\a\b}+R_{\a}{}^{\b}\wedge
{}^{(2)}\lambda^{\a}{}_{\b}\label{GRTel}\,.\ee This can be considered
as a starting point for turning to Lagrangians quadratic in the field
strengths. Amongst the simplest model cases is the {\it purely}
quadratic von der Heyde et al.\cite{vdH2,Ba6} Lagrangian \be V_{\rm
  vdH}=-\frac{1}{2\ell^2}\,\left(T^\a\wedge\vartheta^\b\right)
\wedge{}^*\left(T_\b\wedge\vartheta_\a\right)-\frac{1}{2\kappa}\,
R^{\a\b}\wedge{}^* R_{\a\b}+\frac{1}{2}
\,Q_{\a\b}\wedge\lambda^{\a\b}\,.\label{VvdH}\ee This Lagrangian has
been `derived' by means of the Gordon decomposition
argu-ment\cite{vdH2,Erice}, see also Rumpf\cite{Rumpf}. It may have
problems with the positivity of the energy, see
refs.(\cite{Peter,Kuh,Zhytnikov}), but the situation is not completely
clear to us. The torsion square piece in this Lagrangian, in a
Weitzenb\"ock spacetime, has a classical Newtonian limit.  It differs
from the torsion pieces in the teleparallel Lagrangian (\ref{GRTel})
by a quadratic axial torsion piece (then in (\ref{GRTel}) we had
$-{}^{(3)}T_\a$ instead).
\par

A number of exact classical solutions has been found for the model
(\ref{VvdH}), a Kerr-NUT solution with torsion\cite{Kerr} is amongst
the most prominent ones.  For illustrating the basic features of such
solutions, we display the less complicated subcase, namely the
Baekler-Lee\cite{Baekler,Lee} solution -- this is the
Reissner-Nordstr\"om solution with dynamic torsion -- as a fairly
transparent example. We choose Schwarzschild coordinates
$(t,r,\theta,\phi)$, $M$ = mass, $q$ = electric charge, and find the
orthonormal coframe \bea \vt^{\hat t} &=& {1\over 2}\,[(\Phi
+1)dt+(1-{1\over\Phi})dr]\,,\nonumber \\ \vt^{\hat r} &=& {1\over
  2}\,[(\Phi -1)dt+(1+{1\over\Phi})dr]\,,\nonumber \\ \vt^{\hat
  \th}&=& rd\th \,,\nonumber \\ \vt^{\hat \ph}&=& r\sin\th\,
d\ph\,.\label{BLframe} \eea The corresponding metric reads: \be
ds^2=-\Phi dt^2+{1\over\Phi}dr^2+r^2(d\th^2+\sin^2\th\,
d\ph^2)\label{BLmetric}\,. \ee In the Reissner-Nordstr\"om function
\be \Phi:=1-{{2(Mr-q^2)}\over{r^2}}-{\k\over{4\ell^2}}r^2
\label{BLPhi} \ee the `cosmological term' is induced by the Yang-Mills
type curvature square piece in the Lagrangian (`strong gravity', cf.\ 
ref.(\cite{Yuval3})). If we additionally had a `naked' cosmological
constant, we could shift its value by a suitable choice of the
`strong' coupling constant $\kappa$.  Torsion and curvature are given
by \bea T^{\hat t}=T^{\hat r}&=&{{Mr-{\bf 2}q^2}\over r^3}\; \vt^{\hat
  t}\wedge\vt^{\hat r}\,,\nonumber\\ T^{\hat
  \th}&=&{{Mr-q^2}\over{r^3}}\;\left(\vt^{\hat t}\wedge\vt^{\hat \th}
  -\vt^{\hat r}\wedge\vt^{\hat \th}\right)\,,\nonumber\\ T^{\hat
  \ph}&=&{{Mr-q^2}\over{r^3}}\;\left(\vt^{\hat t}\wedge\vt^{\hat \ph}
  -\vt^{\hat r}\wedge\vt^{\hat \ph}\right)\,,\label{BLtorsion} \eea
and \be
R^{\a\b}={\k\over{4\ell^2}}\,\vt^\a\wedge\vt^\b+{{Mr-q^2}\over{r^2}}\;{}^{(4)}
\RR^{\a\b}\,\label{BLcurv}\,, \ee respectively, where \ba
{}^{(4)}\RR^{{\hat t}{\hat \th}}&=&{}^{(4)}\RR^{{\hat r}{\hat \th}}:=
{\k\over{4\ell^2}}\left( \vt^{\hat t}\wedge\vt^{\hat \th}-\vt^{\hat
    r}\wedge\vt^{\hat \th} \right)\,,\nonumber\\ {}^{(4)}\RR^{{\hat
    t}{\hat \ph}}&=&{}^{(4)}\RR^{{\hat r}{\hat \ph}}:=
{\k\over{4\ell^2}}\left( \vt^{\hat t}\wedge\vt^{\hat \ph}-\vt^{\hat
    r}\wedge\vt^{\hat \ph}\right) \,,\label{BLcurvconst}\ea represent
a tracefree symmetric Ricci piece of the curvature two-form.  The
Coulomb field shows up in the electromagnetic field strength: \be
F=\frac{2\,q}{\ell
  \,r^2}\,\vartheta^{\hat{t}}\wedge\vartheta^{\hat{r}}\,.
\label{BLMax}\ee
\par 
It is a spherically symmetric vacuum solution with Maxwell field,
i.e.\ a Reissner-Nordstr\"om solution with dynamic torsion. We don't
display the solution in its original frame but in a suitably
rotated one\cite{Shira} such that the torsion two-form
(\ref{BLtorsion}) has a `Coulombic' look without global extra factors
in front of the corresponding expressions. Note the factor of {\it
  two} in the $T^{\hat{t}}$ component of (\ref{BLtorsion}) in the
$q^2$-piece.  One interesting feature of this solution is that the
curvature square Lagrangian supplies a constant curvature background
proportional to $\kappa$, see the first piece on the right hand side
of (\ref{BLcurv}). The {\it torsion} is -- this is not an unexpected
feature in the light of our teleparallelism `philosophy' -- induced by
ordinary {\it Newton-Einstein gravity}, as can be read off from
(\ref{BLtorsion}), a fact usually hard to swallow by colleagues who
relate torsion with obscurity. In (\ref{BLtorsion}) torsion is visibly
the translation field strength, a fact made possible by the purely
quadratic torsion piece in the Lagrangian, without a Hilbert-Einstein
type admixture.
\par
The von der Heyde Lagrangian (\ref{VvdH}) is a subcase of the general
quadratic PG Lagrangian\cite{Erice,Go2,Yu3} \ba V_{\rm
  QPG}=\frac{\Lambda}{\ell^2}\,\eta&+&\frac{a_0}{2\ell^2}
\,R^{\a\b}\wedge\eta_{\a\b}+\frac{1}{2\ell^2}\,T^\a\wedge\hodge\left(
  \sum_{M=1}^{3}a_{(M)}\,^{(M)}T_\a\right)\nonumber\\&+&
\frac{1}{2\kappa}\,R^{\a\b} \wedge\hodge\left(
  \sum_{N=1}^{6}b_{(N)}\,^{(N)}R_{\a\b}\right)+\frac{1}{2}
\,Q_{\a\b}\wedge\lambda^{\a\b}\,.\label{QPG}\ea Each of the three
irreducible torsion and six irreducible curvature pieces contributes
to the Lagrangian with an individual weight.  The propagating modes of
this Lagrangian were investigated by Sezgin and van
Nieuwenhuizen\cite{Peter} and by Kuhfuss and Nitsch\cite{Kuh}.  A
subclass of Lagrangians survived their selection criteria motivated by
quantum field theoretical considerations (ghost-freeness, positive
energy). Minkevich\cite{Mink1,Mink2}, amongst others, studied
Friedmann type cosmological models resulting from such a Lagrangian. 

\subsection{Coupling to a scalar field}

It is near at hand to add a dilaton type massless scalar field to this
Lagrangian: \be V_{\Phi{\rm grav}} = V_{\rm QPG}+\frac{1}{2}\,
d\Phi\wedge \hodge d\Phi\,.\label{scalargrav}\ee We found for this
model a remarkable exact solution, a {\it torsion kink}\cite{Ba6}.
Attributing distinctive Higgs-like features to the scalar field, we
arrive at the more general model of Floreanini and Percacci\cite{Fl3}:
\ba\label{VFP} V_{\rm FP}&= & c_1d\Phi\wedge\hodge d\Phi+c_2
d\Phi\wedge\hodge (e_\mu\rfloor T^\mu)+U(\Phi)+\frac{a_0}{2\ell^2}\,
\Phi^2R^{\a\b}\wedge\eta_{\a\b} \\ &+& \frac{1}{2\ell^2}\,
T^\a\wedge\hodge
\left(\sum_{M=1}^{3}a_M\,^{(M)}T_\a\right)+\frac{1}{2}\,
R^{\a\b}\wedge\hodge \left(\sum_{N=1}^{6}b_N\,^{(N)}R_{\a\b}\right)
+\frac{1}{2}\,Q_{\a\b}\wedge\lambda^{\a\b}\,.\nonumber\ea The scalar
$\Phi$ couples to the EC term in a Jordan-Brans-Dicke type way. The
explicit form of the `Higgs' potential $U(\Phi)$ is left open.  Note
the direct coupling $d\Phi\wedge\hodge T$ in (\ref{VFP}) which is,
however, odd in $\Phi$. The authors of (\ref{VFP}), on a quantum field
theoretical level, investigated the renormalizability properties of
their model.
\par
Being relativists, we would be ill-advised if we didn't try to 
give the scalar field $\Phi$ a geometrical meaning, perhaps in 
the context of the Weyl one-form $Q$ which is of the type of a 
gauge potential for dilations anyways. Therefore we lift the last 
Lagrange multiplier and now turn to...
   
\subsection{Metric-affine gauge theory MAG}

Let us, however, first continue the discussion of above of how to
arrive at GR in spite of relaxing the last constraint and liberating
thereby the connection completely from its dominance by the metric.
The naive way, namely just to take a term proportional to $R^{\a\b}
\wedge\eta_{\a\b}$, doesn't work. The projective transformation \be
\Gamma_\a{}^\b\,\longrightarrow\,\Gamma_\a{}^\b+\delta_\a^\b
P\,,\label{proj}\ee with some one-form field $P$, leaves the
Hilbert-Einstein type Lagrangian invariant. Consequently, in such a
model, the connection would only be determined up to a one-form (with
four components). This is unsatisfactory. Moreover, if one coupled the
gauge fields to matter, then only projectively invariant matter
Lagrangians would be allowed, an a priori constraint without physical
justification. Therefore one has to remove the projective invariance
from the gravitational Lagrangian. 
\par
Since a projective transformation changes the trace
$\Gamma_\gamma{}^\gamma$ of a connection and this trace is closely
related to the Weyl one-form,
\begin{equation} \Gamma _{\gamma } {} ^ \gamma = { 2} \,Q + d \ln
  \,\sqrt{|{\rm det}\, g_{\alpha \beta }|}\> ,
\qquad d\Gamma_\gamma{}^\gamma=R_\g{}^\g=2\,dQ\,,
\end{equation}
see Eq.(3.10.13) of ref.(\cite{PRs}), an obvious way to remove the
projective invariance is to add the square of the Weyl
one-form\cite{He10,Dermott}: \be V_{\rm
  GR'}=-\frac{1}{2\ell^2}\,\left(R^{\a\b}\wedge\eta_{\a\b}+
  \beta\,Q\wedge\hodge\, Q\right)\,.\label{GR'} \ee A moment's
reflection will remind ourselves what we have achieved by this
innocently looking Lagrangian (\ref{GR'}): Varying metric, frame, and
connection independently, in vacuum -- that is, in the absence of
matter -- yields the {\it Einstein vacuum equation in a Riemannian
  spacetime.} Generically, if matter is present and supplies
energy-momentum and hypermomentum currents, then the {\it
  hypermomentum}, via the second field equation (\ref{second}), turns
out to be proportional to the {\it post}-Riemannian pieces of the
connection.
\par
The problem with the Lagrangian (\ref{GR'}) is that it has the same
defect as the EC Lagrangian. We have Newton-Einstein gravity and no
further propagating modes. The obvious remedy is to add kinetic terms
and, since we called the Weyl one-form already in earlier, we may want
to make these modes propagating\cite{LordSmalley,Pono}: \be V_{\rm
  GRQ}=-{1\over2\ell^2}\,\left(R^{\alpha\beta}\wedge
  \eta_{\alpha\beta}+ \beta\,Q\wedge\hodge\,Q\right) -{\a\over
  2}\,dQ\wedge\hodge dQ\,.\label{WeylMax} \ee In our first order
formalism the Lagrangian must be expressed in terms of the gauge
potential and their first derivatives. Since $Q$ is already on the
level of a field strength, one may be tempted to forbid the kinetic
$Q$-terms. However, by means of the trace of the zeroth Bianchi
identity, $dQ={1\over 2}R_\g{}^\g$, we can cast out the pseudo-second
order terms and find the well-behaved first order Lagrangian \be
V_{\rm GRQ}=-{1\over2\ell^2}\,\left(R^{\alpha\beta}\wedge
  \eta_{\alpha\beta}+ \beta\,Q\wedge\hodge\,Q\right) -{\a\over
  8}\,R_\alpha{}^\alpha\wedge\hodge R_\b{}^\b \,.\label{WeylMax'}\ee
This might be the {\it simplest} reasonable metric-affine Lagrangian
with propagating dilation modes. Probably it deserves closer
investigation. If one amended (\ref{WeylMax'}) with a piece $\;\sim
R^{\a\b} \wedge\hodge\,{}^{(3)}Z_{\a\b}$, see (\ref{QMA}), then one
would expect the simplest shear modes to arise. Here
$Z_{\a\b}:=R_{(\a\b)}$ is the symmetric (post-Riemannian) and, for
later use, $W_{\a\b}:=R_{[\a\b]}$ the antisymmetric piece of the
curvature two-form.
\par
The trace $R_\b{}^\b$ of the curvature in (\ref{WeylMax'}) represents
an irreducible piece of the curvature, in fact the piece which we
numbered as {\it ten} and called, in our computer algebra
programs\cite{Schru}, DILCURV. Altogether, in a metric-affine space,
the curvature has {\it eleven} irreducible pieces, see
ref.(\cite{PRs}), Table~4.  If we recall that the nonmetricity has
{\it four} irreducible pieces, then the general quadratic Lagrangian
in MAG reads: \ba \label{QMA} V_{\rm QMA} & =&
\frac{1}{2\ell^2}\,\left[-a_0\,R^{\a\b}\wedge\eta_{\a\b}+2\Lambda\,\eta+
  T^\a\wedge\hodge\left(\sum_{I=1}^{3}a_{I}\,^{(I)}T_\a\right)\right.
\nonumber\\&+&\left.
  2\left(\sum_{I=2}^{4}c_{I}\,^{(I)}Q_{\a\b}\right)\wedge\vartheta^\alpha
  \wedge\hodge T^\beta + Q_{\a\b}\wedge\hodge
  \left(\sum_{I=1}^{4}b_{I}\,^{(I)}Q^{\a\b}\right)\right]
\nonumber\\&+ &\frac{1}{2}\,R^{\a\b} \wedge\hodge
\left(\sum_{I=1}^{6}w_{I}\,^{(I)}W_{\alpha\beta} +
  \sum_{I=1}^{5}{z}_{I}\,^{(I)}Z_{\alpha\beta}\right)
\,.\label{general} \ea In spite of this affluence of generality,
Tresguerres\cite{Tr5a}$^,$\footnote{See also
  Tresguerres\cite{Tr4,Tr5,Tr5b,Tr5c}, Mac\'{\i}as et
  al.\cite{Alfredo1}, and, for cosmological models,
  Minkevich\cite{Mink3}.}$\,$ was able to find two exact solutions of
the corresponding vacuum field equations in a most remarkable piece of
work. He also discussed\cite{Tr5a} the reasons for introducing the
mixed $Q\wedge\vartheta\wedge\hodge\, T$ term. Generically Tresguerres
found a Baekler-Lee type solution with torsion, see (\ref{BLframe}) to
(\ref{BLMax}) above, but the {\it dilaton charge takes the place of
  the electric charge.} Thus the Weyl one-form $Q$, the
quasiMaxwellian potential, has a $1/r$-behavior and is closely
interwoven with the torsion vector. In the second solution, the metric
is the same, but, additionally, two types of {\it shear charges}
emerge. The Tresguerres solutions with dilation and shear charges are
something qualitatively new in gauge models of gravity.
\par
Tucker and Wang\cite{Tucker1} put $\b=0$ in (\ref{WeylMax}) or
(\ref{WeylMax'}), \be V_{\rm
  TW}=-{1\over2\ell^2}\,R^{\alpha\beta}\wedge \eta_{\alpha\beta}
-{\a\over 2}\,dQ\wedge\hodge
dQ=-{1\over2\ell^2}\,R^{\alpha\beta}\wedge \eta_{\alpha\beta}
-{\a\over 8}\,R_\alpha{}^\alpha\wedge\hodge R_\b{}^\b
\,,\label{TWLagr} \ee i.e.\ they excluded the massive term of the Weyl
field from further consideration. Rather, they explored the analogy of
the Weyl one-form $Q$ to the Maxwell potential $A$ under these
circumstances. Their field equations can be read off from
(\ref{first}), (\ref{second}), and (\ref{gaugeem}) straightforwardly
as, \ba -e_\a\rfloor V_{\rm TW}+(e_\a\rfloor R_\b{}^\g)\wedge{{\6
    V_{\rm TW}}\over{\6 R_\b{}^\g}}&=&\Sigma_\a\,,\\ -D\left({\partial
    V_{\rm TW}\over\partial R_\a{}^\b}\right)&=&\Delta^\a{}_\b\,,\ea
or \bea {1\over 2}\,\eta_{\a\b\g}\wedge
R^{\b\g}+{\a\over{8}}\left[(e_\a\rfloor R_\b{}^\b )\wedge{}^*R_\g{}^\g
  -(e_\a\rfloor{}^*R_\b{}^\b)\wedge R_\g{}^\g\right] &=&
\ell^2\,\Sigma_\a\,, \label{TW1} \\ {1\over 2}\,D\eta^\a{}_\b
-{\a\over 4}\,\d^\a_\b\, d\,\hodge R_\g{}^\g &=&
\ell^2\,\Delta^\a{}_\b\,.
\label{TW2}
\eea Tucker and Wang\cite{Tucker1} found Baekler-Lee type vacuum
solutions with dilation (`Weyl') charge, just as Tresguerres, but,
in addition, they presented a highly interesting solution with propagating
massless spinor matter as source.
\par
Nevertheless, one should recognize that the Lagrangian (\ref{TWLagr})
is not without problems. Because the massive piece $\sim
Q\wedge\hodge\, Q$ is missing, the gauge Lagrangian $V_{\rm TW}$ is
invariant under the special projective transformation \be
\Gamma_\a{}^\b\,\longrightarrow\,\Gamma_\a{}^\b+\delta_\a^\b
\,d\,p\,.\label{proj'} \ee Whereas such a type of `gauge'
transformation is desirable for an internal $U(1)$-connection -- like
in Maxwell's theory -- it is definitely dangerous in the context of a
dilation transformation. The linear connection is only determined up
to the transformation (\ref{proj'}), that is, not all of the 64
components are uniquely determined in the TW-model, see also the
analysis of Teyssandier and Tucker\cite{Tucker2}.
\par
The case studies of Tresguerres and Tucker-Wang taught us that the
Weyl one-form plays a particular role in representing a dilation type
field that ought to be useful in the context of the breaking of the
dilation symmetry. Therefore there were attempts by Mielke et al., see
ref.(\cite{PRs}) Sec.\ 6, to superimpose on the Lagrangian
(\ref{QMA}) an additional conformal symmetry in order to have a
massless theory, free of dimensionful coupling constants at the
beginning. Then one couples to a hypothetical dilaton field $\sigma$,
\begin{equation} V =-\frac{\sigma^{2}}{2}\, R^{\alpha\beta}
  \wedge\eta_{\a\b} + {1\over 2}\,(D\sigma)\wedge\hodge
  D\sigma+{\lambda\over 4}\,\sigma^ {4}\,\eta \,,\end{equation} and
breaks the dilation symmetry, thereby arriving at a low energy massive
Lagrangian. Analogous approaches were proposed by Gregorash and
Papini\cite{Gr3,Gr3a}, Hochberg and Plunien\cite{Hochberg}, and by
Poberii\cite{Poberii}. Recently Paw{\l}owski and
R{\c{a}}czka\cite{PW,PW2,Tsantilis} developed similar models in a
Riemannian spacetime, though. Nevertheless, they required conformal
invariance for the Lagrangian they started from, coupled to a dilation
field etc.. There are close relationships between the gravitational
sectors of these models which lead us to the belief that the P\&R
model should be redrafted in the framework of a Weyl-Cartan, if not of
a metric-affine spacetime.
\par
With all these developments from different quarters, we have here --
perhaps for the first time -- a consistent framework for gauge
models carrying both, non-trivial torsion {\it and} nonmetricity.

\section{Acknowledgments}
One of the authors would like to thank Venzo de Sabbata, Peter
Bergmann, and H.-J.\ Treder for the invitation to lecture at this
Erice School and his coauthors\cite{PRs} Dermott McCrea$^\dagger$
(Dublin), Eckehard Mielke (Kiel/Mexico City), and Yuval Ne'eman (Tel
Aviv/Austin) for great help.  Moreover, he is very obliged to H.A.
Kastrup (Aachen), G.  S\"ussmann (Munich), Ryszard R\c{a}czka
(Warsaw), and H. Petry (Bonn) for seminar invitations where different
versions of these lectures were given. We are most grateful to our
colleagues David Hartley (St.Augustin), Yuri Obukhov (Moscow/Cologne),
and Romualdo Tresguerres (Madrid) for help and many clarifying
discussions.

\section{References}

\end{document}


\end{thebibliography}

\end{document}